\documentclass[%
 reprint,
superscriptaddress,
 amsmath,amssymb,
 prd,
]{revtex4-1}

\usepackage[utf8]{inputenc}
\usepackage{amsmath}
\usepackage{caption}
\usepackage{ulem}

\usepackage{graphicx}
\usepackage{hyperref}
\usepackage[labelformat=simple]{subcaption}

\usepackage{xcolor}
\usepackage{orcidlink}
\definecolor{orange}{rgb}{1.0, 0.5, 0.0}

\newcommand{\orcid}[1]{\href{https://orcid.org/#1}{\textcolor{blue}{\aiOrcid}}}

\hypersetup{
    colorlinks=true,
    linkcolor=blue,
    filecolor=magenta,      
    urlcolor=cyan
    }
\begin{document}

\title{The Influence of Muons, Pions, and Trapped Neutrinos on Neutron Star Mergers}

\author{Michael A. Pajkos\orcidlink{0000-0002-4983-4589}}
\email[E-mail: ]{mpajkos@caltech.edu}

\affiliation{TAPIR, Mailcode 350-17, California Institute of Technology, Pasadena, CA 91125, USA}

\author{Elias R. Most\orcidlink{0000-0002-0491-1210}}%
\affiliation{TAPIR, Mailcode 350-17, California Institute of Technology, Pasadena, CA 91125, USA}
\affiliation{Walter Burke Institute for Theoretical Physics, California Institute of Technology, Pasadena, CA 91125, USA}

\date{\today}

\begin{abstract}
 The merger of two neutron stars probes dense matter in a hot, neutrino-trapped regime. In this work, we investigate how fully accounting for pions ($\pi$), muons ($\mu$), and muon-type neutrinos ($\nu_\mu$) in the trapped regime may affect the outcome of the merger. 
 By performing fully general-relativistic hydrodynamics simulations of merging neutron stars with equations of state to which we systematically add those different particle species, we aim to provide a detailed assessment of the impact of muons and pions on the merger and post-merger phase.
 In particular, we investigate the merger thermodynamics, mass ejection and gravitational wave emission. Our findings are consistent with previous expectations, that the inclusion of such microphysical degrees of freedom and finite temperature corrections leads to frequency shifts on the order of $100-200\, \rm Hz$ in the post-merger gravitational wave signal, relative to a fiducial cold nucleonic equation of state model.  
\end{abstract}

\maketitle

\section{Introduction}
\label{intro}
Neutron star mergers are ideal probes for nuclear matter under extreme conditions \cite{Lattimer:2006xb,Lattimer:2015nhk,Oertel:2016bki,MUSES:2023hyz}. Neutron star interiors can reach densities beyond several times nuclear saturation, at which physics beyond those of neutrons (n) and protons (p) can be probed \cite{Ozel:2016oaf,Oertel:2016bki}. 
These can include exotic degrees of freedom such as hyperons \cite{Schaffner:1995th,Tolos:2020aln} or deconfined quarks \cite{Weber:2004kj,Annala:2019puf}, but also mesons, including pions ($\pi^{\pm}$) \cite{Baym:1974vzp,Haensel:1982zz,Brandt:2018bwq}. Revealing this complex and intricate interplay of the physics of the strong interaction is at the core of modern astrophysical neutron star research. As such, various ways have been proposed and investigated to reveal matter under extreme conditions in neutron stars. These range from X-ray observations of neutron star cooling curves and hotspots on neutron stars \cite{Ozel:2016oaf,Watts:2016uzu}, which has yielded novel constraints on the dense matter equation of state \cite{Raaijmakers:2019dks,Bogdanov:2021yip,Miller:2021qha,Vinciguerra:2023qxq,Rutherford:2024srk}, especially when combined with advances in chiral effective field theory descriptions of nuclear matter around saturation \cite{Drischler:2021kxf}.

Within the realm of gravitational wave astrophysics, detections of gravitational waves from merging neutron stars \cite{LIGOScientific:2017vwq,LIGOScientific:2020aai} have the potential to directly probe and constrain dense matter physics, including finite temperature effects \cite{Baiotti:2019sew}. The inspiral part of the gravitational waveform can constrain neutron star radii and the cold equation of state \cite{Cutler:1994ys,Flanagan:2007ix}, with remarkable constraints obtained from the first neutron star merger event GW170817 (e.g., \cite{Chatziioannou:2018vzf,Raithel:2018ncd,Annala:2017llu,Most:2018hfd,LIGOScientific:2018cki}, see also \cite{Bauswein:2017vtn,Margalit:2017dij,Rezzolla:2017aly,Ruiz:2017due,Shibata:2019ctb,Nathanail:2021tay,Tan:2020ics,Most:2020bba,Fattoyev:2020cws}). 
With next-generation facilities, future detections of post-merger gravitational waves have the potential to detect the kilohertz oscillations of the neutron-star merger remnant formed in intermediate and low mass mergers \cite{Cutler:1992tc,Shibata:2005xz}. These frequencies are quasi-universally related to properties of the dense matter equation of state \cite{Bauswein:2011tp,Bauswein:2012ya,Takami:2014tva,Takami:2014zpa,Bernuzzi:2014owa,Rezzolla:2016nxn,Vretinaris:2019spn,Breschi:2022xnc,Topolski:2023ojc} (see also \cite{Raithel:2022orm} for bounds on this universality). The post-merger phase probes higher densities and temperatures than present in the inspiral, and could reveal the appearance of quark matter \cite{Most:2018eaw,Bauswein:2018bma,Most:2019onn,Weih:2019xvw,Liebling:2020dhf,Prakash:2021wpz,Huang:2022mqp,Ujevic:2022nkr}, hot dense matter \cite{Bauswein:2010dn,Perego:2019adq,Figura:2020fkj,Raithel:2021hye,Fields:2023bhs,Raithel:2023gct,Raithel:2023zml,Villa-Ortega:2023cps,Miravet-Tenes:2024vba}, hyperons \cite{Sekiguchi:2011mc,Radice:2016rys,Blacker:2023opp} and neutrinos \cite{Alford:2017rxf,Most:2021zvc,Zappa:2022rpd,Espino:2023dei,Most:2022yhe}, fueling effective chemical reactions inside the merger remnant. While the latter could in principle affect the outcome of the merger via an effective viscosity \cite{Most:2021zvc}, the strength of the effect depends strongly on the neutrino conditions and opacity \cite{Espino:2023dei,Most:2022yhe}. In particular, it has been found that likely the remnant will be optically thick to neutrino emission, leading to an effective trapping of neutrinos and to a correction of the equation of state \cite{Espino:2023dei}.
This neutrino trapped regime can depend crucially on the microphysical interactions included \cite{Alford:2021lpp}, including through the appearance of muons ($\mu^\pm$) 
and pions, $\pi^\pm$ \cite{Harris:2024ssp} (see also \cite{Fore:2019wib}).
Recent numerical studies have further been aimed at quantifying the impact of these particles independently \cite{loffredo:2023,vijayan:2023}.
Similar conclusions have also been found for core-collapse supernovae, where the inclusion of muons and pions may critically affect the explosion mechanism \cite{Bollig:2017lki,Guo:2020tgx,Fischer:2020vie}.\\
Recent works in the field have made substantial progress towards quantifying the effects of additional particles. Ref. \cite{vijayan:2023} varies the mass of the pion and quantifies the characteristics of the NSNS dynamics and resulting GW signal.  Ref. \cite{loffredo:2023} uses an advanced postprocessing scheme to quantify the changes in pressure in the presence of muons and trapped neutrinos, finding changes in pressure of order $10\%$.  More recently, Ref. \cite{gieg:2024} allows for the advection of muons, paired with a neutrino leakage scheme, while Ref.  \cite{ng:2024} performs 5 species neutrino transport alongside muonic reactions in NSNS.\\

Building upon these works, we perform fully general-relativistic neutron star merger simulations,  which model the influence of mesons, such as pions ($\pi^\pm$) in their thermal and condensed state, and leptons, such as muons ($\mu^\pm$) and (anti)neutrinos ($\nu$) in the fully trapped regime. We then analyze the impact on the merger dynamics, gravitational wave emission, and mass ejection from the system. \\

Our work is organized as follows, in Section \ref{sec:particle_processes} we outline the relevant particle processes we choose to model.  Section \ref{sec:particle_stats} provides the statistic description of the various particles. Section \ref{sec:eos_update} we outline the relevant steps to update nuclear EOS to include these particles for the NSNS system, with more detail in Appendix \ref{app:eos_steps}.  Section \ref{sec:thermo_qts} provides thermodynamic contributions from each species.  Section \ref{sec:eos_models} introduces the EOSs used in this work.  Section \ref{sec:eos_properties} describes the general properties of the newly constructed EOSs. Section \ref{sec:tov} quantifies the impact of different species on isolated NSs.  Section \ref{sec:num_methods} details the numerical tools in this work.  Section \ref{sec:sim_results} describes our simulation results.  Lastly, Section \ref{sec:conclusions} concludes.

Unless otherwise noted, in this work we adopt units $G=c=1$, where $G$ is the gravitational constant, and $c$ the speed of light.



\section{Methods}

In the following, we outline the construction of the equation of state, as well as the setup for our numerical simulations.

\subsection{Particle Processes}
\label{sec:particle_processes}

We begin by discussing the main particle interactions of muons and pions relevant to neutron star mergers. Our presentation largely follows that of Refs. \cite{loffredo:2023,vijayan:2023}.\\

The principal decay channel of a charged pion ($\pi^\pm$, with bare mass $m_\pi \sim 140$ MeV) into charged muons, $\mu^\pm$, is as follows,
\begin{equation}
    \pi^- \leftrightarrow \mu^- + \bar{\nu}_\mu\,,
\end{equation}
where $\bar{\nu}_\mu$ denotes a muon anti-neutrino.
Charged muons (with bare mass $m_\mu \sim 106$ MeV) will further decay into electrons, $e^-$, electron type antineutrino, $\bar{\nu}_e$, and muon type neutrino, ${\nu}_\mu$,
\begin{equation}
    \mu^- \leftrightarrow e^- + \bar{\nu}_e + \nu_\mu.
    \label{eq:mu_decay}
\end{equation}
The last relevant reaction is neutron ($n$, with bare mass $m_n \sim 940$ MeV) decay into a proton ($p$), electron, and electron type antineutrino
\begin{equation}
    n \leftrightarrow p + e^- + \bar{\nu}_e.
    \label{eq:n_decay}
\end{equation}
We now assume that neutrinos inside the neutron star merger remnant are trapped \cite{Espino:2023dei}, and therefore approximate weak decays inside the hot and dense merger remnant \cite{Perego:2019adq}, as being in weak-interaction equilibrium. As a result, we can equate the reactions in terms of their chemical potentials, $\mu_i$,
\begin{equation}
    \mu_{\pi^\mp} = \mu_{\mu^\mp} -\mu_{\nu_\mu}= \mu_{e^\mp}  - \mu_{\nu_e}= \pm(\mu_n - \mu_p)\,,
    \label{eq:chem}
\end{equation}
where the different signs correspond to negatively and positively charged current reactions, respectively.
Apart from weak-interaction equilibrium, we also need to account for charge neutrality. As a result, the
overall particle fractions, $Y_i= n_i/ n_b$, where $n_i$ are the particle number densities and $n_b$ is the baryon number density, obey
\begin{equation}
    Y_p = Y_e + Y_\pi + Y_\mu,
    \label{eq:charge}
\end{equation}
where $Y_e = Y_{e^-} - Y_{e^+}$,  $Y_\pi = Y_{\pi^-} - Y_{\pi^+} + Y_\pi^c$, and $Y_\mu = Y_{\mu^-} - Y_{\mu^+}$.  
Here $Y_\pi^c$ accounts for the presence of a negatively charged pion condensate.  Particles that follow Bose-Einstein statistics can form multiple particles in the same quantum state at the lowest energy level of the system.  These particles have rest mass, but no kinetic energy, and do not contribute pressure to the surrounding system.  As such, we call pions which do not condense ($Y_{\pi^-}, Y_{\pi^+}$) {\it thermal} pions.  Neutral pions ($\pi_0$) are added by assuming a vanishing chemical potential, $\mu_{\pi_0}=0$.

Similar to Ref. \cite{loffredo:2023}, we define the lepton fraction for a species $i$, described as 
\begin{equation}
    Y_{l,i} = (Y_{i^-} - Y_{i^+}) + Y_{\nu_i} - Y_{\bar{\nu}_i} \,|\,  i \in \{e, \mu\}.
\end{equation}
One central assumption to our hydrodynamic scheme is that the electron lepton number in the simulation is advected along with the fluid in the trapped regime \cite{Perego:2019adq,Endrizzi:2019trv},
\begin{align}
    u^\mu \nabla_\mu Y_{l,e} = 0\,,
\end{align}
where $u^\mu$ is the fluid four-velocity, which is in good agreement with simulations of neutron star mergers with more advanced treatments of neutrino transport \cite{Espino:2023dei}.  $Y_{l,\mu}$ is chosen to be a constant of 0.01 below a density threshold of $10^{14}$g cm$^{-3}$.  Above this threshold, it is parameterized as a function of density $Y_{l,\mu}(\rho)$.  For a detailed explanation of our choice of $Y_{l,\mu}$, see Appendix \ref{app:eos_steps}.  We approximate the particle masses in this work with corresponding vacuum rest masses $m_\mu = 105.70$ MeV, $m_{\pi^\pm} = 139.57$ MeV, and $m_{\pi^0} = 134.98$ MeV.  In reality, varying masses of the pions in matter \citep{fore:2024,vijayan:2023} will increase, producing a less pronounced effect of pions on the dynamics, allowing increased muon effects.  Likewise, effective muon masses can change due to relativistic effects \citep{shternin:2008}.

\subsection{Statistical Descriptions of Pions and Muons}
\label{sec:particle_stats}

In order to describe muon and pion corrections to the equation of state, we need to translate Eqs. \eqref{eq:chem} and \eqref{eq:charge} into corrections of the energy density and pressure of the equation of state. As an intermediate step, this involves computing the particle fractions of $e^\pm$, $\mu^\pm$, $\pi^\pm$, $\nu_i$, and $\bar{\nu}_i$.  We will also provide the expressions for the number density, $n_i$, of the respective particles.

\subsubsection{Muons and Electrons}

Leptons ($l = \mu,e$) are treated as an ideal Fermi gas whose number density can be described as (Equation (16) of \cite{loffredo:2023}),

\begin{equation}
    n_{l^\pm} = K_l \theta_l^{3/2}\Big[F_{1/2}(\eta'_{l^\pm}, \theta_l) + \theta_l F_{3/2}(\eta'_{l^\pm}, \theta_l)\Big],
    \label{eq:n_fermi}
\end{equation}
for $K_l = 8 \sqrt{2}\pi(m_l c^2 / hc)^3$
, $\theta_l = k_\mathrm{B}T / (m_l c^2)$, $\eta'_{l^\pm} = (\mu_{l^\pm} - m_{l} c^2)/(k_\mathrm{B}T)$, $F_k$ are Fermi functions of order $k$, $m_l$ is the mass of the lepton, and $h$ and $k_B$ are Planck's and the Boltzmann constant, respectively.  These Fermi functions are defined as (Equation (6) of \cite{timmes:1999})
\begin{equation}
    F_k(\eta,\theta) = \int_0^\infty dx \frac{x^k\sqrt{1 + 0.5 \theta x}}{e^{x - \eta} + 1}.
    \label{eq:fermi}
\end{equation}

\subsubsection{Pions}

Pions, by contrast, are treated as a free Bose gas.  Note, the use of $\mu$ in this section refers to chemical potential, rather than muons.  As a Bose gas, the pion number density is given by \footnote{Equation (3.11.4) of \url{https://www.pas.rochester.edu/~stte/phy418S21/units/unit_3-11.pdf}},

\begin{equation}
    n_\pi = \frac{1}{\lambda^3}g_{3/2}(z) = \frac{1}{\lambda^3}\frac{2}{\sqrt{\pi}}\int_0^\infty dy \frac{y^{1/2}}{z^{-1}e^y - 1}\,,
    \label{eq:n_bose}
\end{equation}

where $\lambda = (h^2 / 2\pi m k_\mathrm{B}T)^{1/2}$ and the fugacity is $z = e^{\mu/(k_\mathrm{B}T)} = e^{(\tilde{\mu} - m)/(k_\mathrm{B}T)}$; for the corresponding antiparticle, $z = e^{(-\tilde{\mu} - m)/(k_\mathrm{B}T)}$.  The integral in Equation (\ref{eq:n_bose}) can be approximated with an infinite series, simplifying to
\begin{equation}
    n_\pi =\frac{1}{\lambda^3}\Bigg(z + \frac{z^2}{2^{3/2}} + \frac{z^3}{3^{3/2}} + ... \Bigg).
\end{equation}
We follow the nomenclature of \cite{thorne:2017} where $\tilde{\mu}$ is the chemical potential of the particle with the rest mass included, and $\mu$ is the chemical potential without the particle rest mass; concretely, $\mu_i = \tilde{\mu_i} - m_i$.\\
As an important note for pions, if $\tilde{\mu} > m_{\pi^-}$ a negatively charged pion condensate ($\pi_c^-$) will form.  In regions of our simulation where this is the case, to populate the \textit{thermal} pions, $\pi^-$ one follows the above Bose-Einstein statistics, with a chemical potential of the thermal pion equal to the charged pion rest mass, $\tilde{\mu}_{\pi^-} = m_{\pi^-}$.  To calculate the number density of the condensate, one must follow a more detailed procedure in Appendix \ref{app:eos_steps}, that relies on balancing charge neutrality.  

\subsubsection{Neutrinos}
In this work, we assume that neutrinos are trapped through the neutron star merger remnant, which is consistent with recent simulations using full neutrino transport \cite{Espino:2023dei}. In the neutrino trapped regime, neutrinos behave as a massless Fermi gas, whose number density can be described as \cite{loffredo:2023},
\begin{equation}
    n_\nu = \frac{4\pi}{(hc)^3}(k_\mathrm{B}T)^3 F_2(\eta_\nu)\exp(-\rho_\mathrm{lim} / \rho)
    \label{eq:nu_number}
\end{equation}
where $\exp(-\rho_\mathrm{lim} / \rho)$ is an exponential damping factor \cite{kaplan:2014} to model trapped neutrinos above $\rho_\mathrm{lim} = 10^{14}$ g cm$^{-3}$.  Here, the degeneracy parameter of the neutrinos is calculated as 
\begin{equation}
    \eta_{\nu_l} = (\mu_p - \mu_n + \mu_l) / k_\mathrm{k_\mathrm{B}} T
\end{equation}
and $\eta_{\bar{\nu}} = -\eta_\nu$.
In practice, these enter our calculation via the net number fraction of neutrinos, represented by $Y_\nu - Y_{\bar{\nu}} \propto n_\nu - n_{\bar{\nu}}$.  Applying Equation (\ref{eq:nu_number}), we use the exact expression provided by \cite{bludman:1978}

\begin{equation}
Y_\nu - Y_{\bar{\nu}} \propto F_2(\eta_\nu) - F_2(-\eta_\nu) = \frac{1}{3}\eta_\nu (\pi^2 + \eta_\nu^2)
\end{equation}
which provides a more accurate expression of the net number density, without numerical integration of Fermi integrals $F_2(\eta_\nu)$.  Outside the trapped regime, neutrino emission is modelled with a leakage scheme, whose details are specified in Section \ref{sec:num_methods}.

\subsection{Equation of State}
\label{sec:eos_update}

Armed with the statistical descriptions for the number density for each species, we now outline our methodology to populate a given equation of state with each new particle.  Begin with charge conservation for the proton fraction

\begin{equation}
    Y_p = Y_e + Y_\mu + Y_\pi.
\end{equation}

Likewise, we assert lepton number conservation for species of electrons and muons

\begin{equation}
    Y_{l,e} = Y_e + Y_{\nu_e} - Y_{\bar{\nu}_e}
    \label{eq:e_lepton}
\end{equation}
and
\begin{equation}
    Y_{l,\mu} = Y_\mu + Y_{\nu_\mu} - Y_{\bar{\nu}_\mu}.
    \label{eq:mu_lepton}
\end{equation}

Combining the previous three equations we yield,

\begin{equation}
    \hat{Y}_p = Y_{l,e} - (Y_{\nu_e} - Y_{\bar{\nu}_e}) + Y_{l,\mu} - (Y_{\nu_\mu} - Y_{\bar{\nu}_\mu}) + Y_\pi.
    \label{eq:charge_trial}
\end{equation}
Here $\hat{Y}_p$ indicates our iteration variable of choice that will eventually converge to $Y_\mathrm{p,new}$ for our updated equation of state.

Equation (\ref{eq:charge_trial}) provides a modular framework to add particles to the system.  In our work, we create three new variants by modifying the {\it base} equation of state. The first variant purely accounts for electron type (anti)neutrinos, the $+\nu_e$ case.  The $+\nu_e$ case zeros out the $Y_{l,\mu}$, $(Y_{\nu_\mu} - Y_{\bar{\nu}_\mu})$, and $Y_\pi$ terms.  The second variant accounts for electron type (anti)neutrinos and pions, the $+\nu_e + \pi$ case.  This variant zeros out the $Y_{l,\mu}$ and $(Y_{\nu_\mu} - Y_{\bar{\nu}_\mu})$ terms. The third variant accounts for electron type (anti)neutrinos, pions, muons, and muon type (anti)neutrinos, the $+ \nu_{e,\mu} + \pi + \mu$ case.  This variant includes all terms in Equation (\ref{eq:charge_trial}).  In Appendix \ref{app:eos_steps}, we enumerate our procedure, in detail, to add the new particle species to the EOS.

\subsection{Calculating Thermodynamic Quantities}
\label{sec:thermo_qts}

Having calculated the charge fractions (or number densities) of pions and muons, we can calculate the pressure, $P$, and specific internal energy, $\epsilon$, of the pions, muons, and trapped neutrinos.  In the following sections, we list the expressions for energy density, that has units of energy per volume.  To convert the energy density expressions to specific internal energy, simply divide by the rest mass density $\epsilon_{\mu/\pi} = \varepsilon_{\mu/\pi} / \rho$.
\subsubsection{Pions}

For pions, we leverage Bose-Einstein statistics. 
 For the pressure, 
 
\begin{equation}
    P_\mathrm{Bose} = g_{5/2}(z(T))\Bigg(\frac{2\pi m_\pi}{h^2}\Bigg)^{3/2}(k_\mathrm{B} T)^{5/2},
\end{equation}
where $g_{5/2}(z(T))$ can be described as
\begin{equation}
    g_{5/2}(z(T)) = \frac{1}{\Gamma(5/2)}\int_0^\infty dy \frac{y^{3/2}}{z^{-1}e^y - 1}.
\end{equation}
Note, $\Gamma(5/2) = 3\sqrt{\pi}/4$ and $g_{5/2}(z=1) = \zeta(5/2) \sim 1.342$.
For the energy density \footnote{Equation (3.8.19) of \url{https://www.pas.rochester.edu/~stte/phy418S21/units/unit_3-11.pdf}}, $\varepsilon_\mathrm{Bose} = \frac{3}{2}p_\mathrm{Bose} + n_\pi m_\pi c^2$.  Note the additional contributions from the rest mass, $n_\mu m_\mu c^2$.




\subsubsection{Muons}
\label{ssec:FD_prims}
For muons, we leverage Fermi-Dirac statistics \cite{bludman:1977,loffredo:2023}.  For pressure,

\begin{equation}
    P_{\mu^\pm} = \frac{1}{3}K_\mu m_\mu c^2 \theta_\mu^{5/2}\Big[2F_{3/2}(\eta'_{\mu^\pm}, \theta_\mu) + \theta_\mu F_{5/2}(\eta'_{\mu^\pm}, \theta_\mu)\Big].
\end{equation}
For the energy density, 

\begin{equation}
    \begin{split}
            \varepsilon_{\mu^\pm} &= K_\mu m_\mu c^2 \theta_\mu^{5/2}\Big[F_{3/2}(\eta'_{\mu^\pm}, \theta_\mu) + \theta_\mu F_{5/2}(\eta'_{\mu^\pm}, \theta_\mu)\Big] \\
    &+ n_\mu m_\mu c^2.
    \end{split}
\end{equation}
Note, the last term in the equation accounts for contributions from the rest mass of the muons.


\subsubsection{Neutrinos}

We treat neutrinos as a massless Fermi gas \cite{loffredo:2023}.  For the energy density,

\begin{equation}
    \varepsilon_\nu = \frac{4\pi}{(hc)^3}(k_\mathrm{B}T)^4 F_3(\eta_\nu) \exp(-\rho_\mathrm{lim} / \rho)
\end{equation}
where the term $\exp(-\rho_\mathrm{lim} / \rho)$ is used to smoothly cutoff neutrino trapping above $10^{14}$ g cm$^{-3}$ \cite{perego:2019}.  Note, in practice, the calculations of $F_3(\eta_\nu)$ can be cumbersome.  However, since we are interested in the contributions from antineutrinos as well, the expressions for the net energy density from neutrinos and corresponding antineutrinos becomes 

\begin{equation}
    \varepsilon_\nu + \varepsilon_{\bar{\nu}} \propto F_3(\eta_\nu) + F_3(-\eta_\nu) = \frac{7\pi^4}{60} + \frac{1}{2}\eta^2(\pi^2 + \frac{1}{2}\eta^2),
    \label{eq:eps_nu}
\end{equation}
where in the last expression, we make use of the properties of sums of Fermi integrals \cite{bludman:1978}.  This expression both simplifies computation and is more precise than performing numerical integration.  The expression for pressure simply follows from that of an ultrarelativistic gas,
\begin{equation}
    P_\nu = \varepsilon_\nu / 3.
\end{equation}

\subsection{Equation of State Models}
\label{sec:eos_models}

In order to apply pion and muon corrections, we need to adopt an underlying equation of state framework. We here adopt two models, SFHo \cite{Steiner:2012rk} and DD2 \cite{hempel:2009,typel:2005ba,typel:2009sy} EOSs. The calculations used in both EOSs are based on a relativistic mean field model for nucleons.  Both unmodified tables tabulate basic thermodynamic quantities (e.g., pressure or speed of sound) against three independent variables $(\rho, T, Y_p)$.  These tables assume nuclear statistical equilibrium (NSE) among the constituent particles: nuclei, nucleons, electrons, positrons, and photons.  For these unmodified tables, only the protons, electrons, and positrons are assumed to contribute to the charge fraction, or $Y_p = Y_{e^-} - Y_{e^+}$.  As outlined in the Appendix \ref{app:eos_steps}, we detail our procedure to create tabulated EOSs against $(\rho, T, Y_{l,e})$.

Corrections for pions and muons are applied to the entirety of the new EOS table, whereas EOS contributions from trapped neutrinos are only added in the regime roughly above the neutrino trapping limit of $10^{14}$ g cm$^{-3}$ due to the exponential damping factor of $\exp(-\rho / \rho_\mathrm{lim})$ in Equation (\ref{eq:eps_nu}).  Ideally, this approximation to neutrino trapping should not be used during the inspiral phase of the merger because there are no expected neutrino contributions from isolated NS companions.  Because we do not see major neutrino fractions in the isolated companions, and for the simplicity of a single EOS table during the entire evolution, we employ this approximation.







\begin{figure*}
    \centering
    \includegraphics{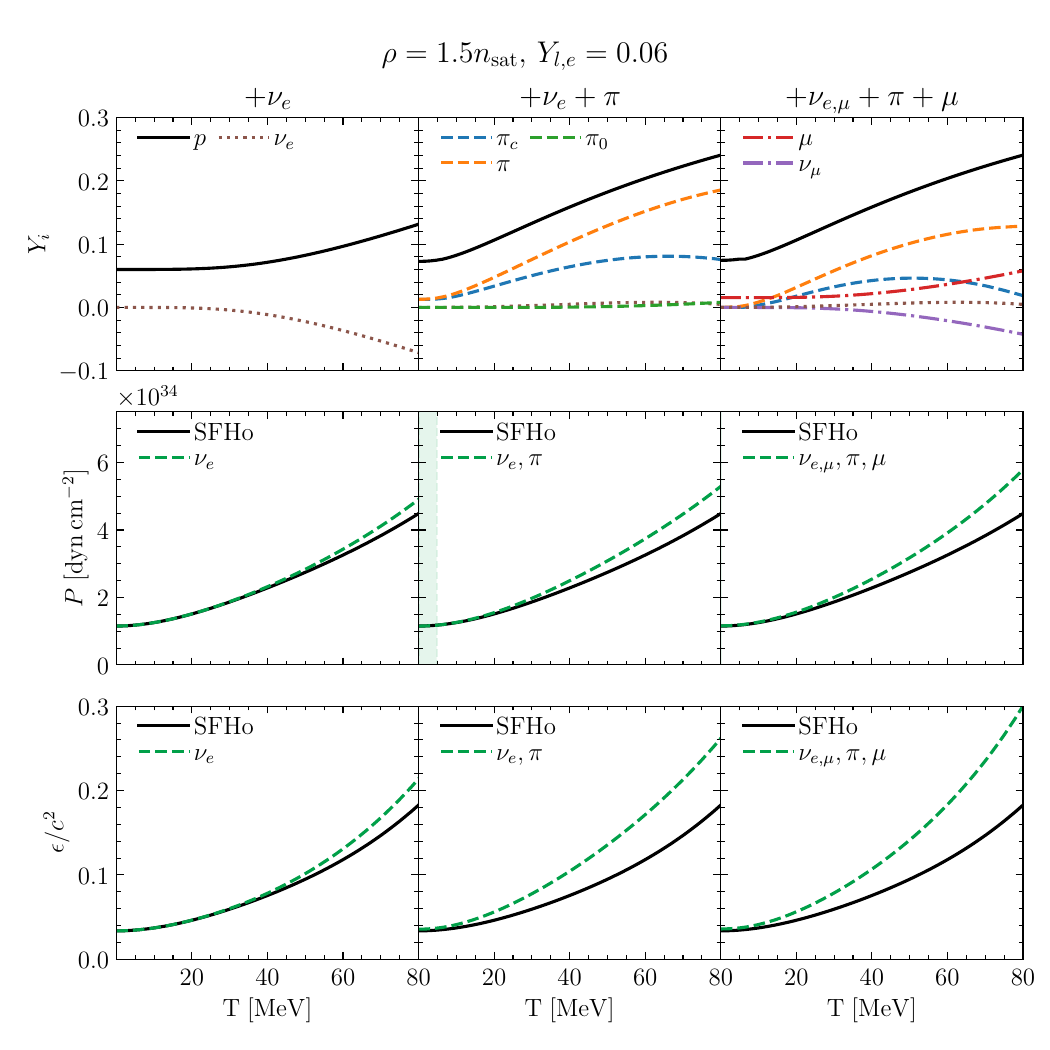}
    \caption{
    Physical quantities for the SFHo EOS, taken at rest-mass density $\rho = 3.9\times 10^{14}$ g cm$^{-3}$ and lepton fraction $Y_{l,e} = 0.06$.  The left column represents only electron type (anti)neutrinos $(\nu_e)$.  The middle column represents $\nu_e$ and pions $(\pi)$ (including condensed pions, $\pi_c$). The right column includes $\nu_e$, $\pi$, muons $\mu$, and muon (anti)neutrinos $(\nu_\mu)$.  (Top row) Particle fractions for various species.  (Middle row) Pressure, $P$, as a function of temperature, $T$, for the unmodified SFHo equation of state and SFHo with additional particles.   The shaded region represents where the pressure $P_\mathrm{SFHo} > P_{+\nu_e +\pi}$.  (Bottom row) Specific internal energy, $\epsilon$, as a function of temperature.}
    \label{fig:9_panel_SFHo}
\end{figure*}

\begin{figure*}
    \centering
    \includegraphics{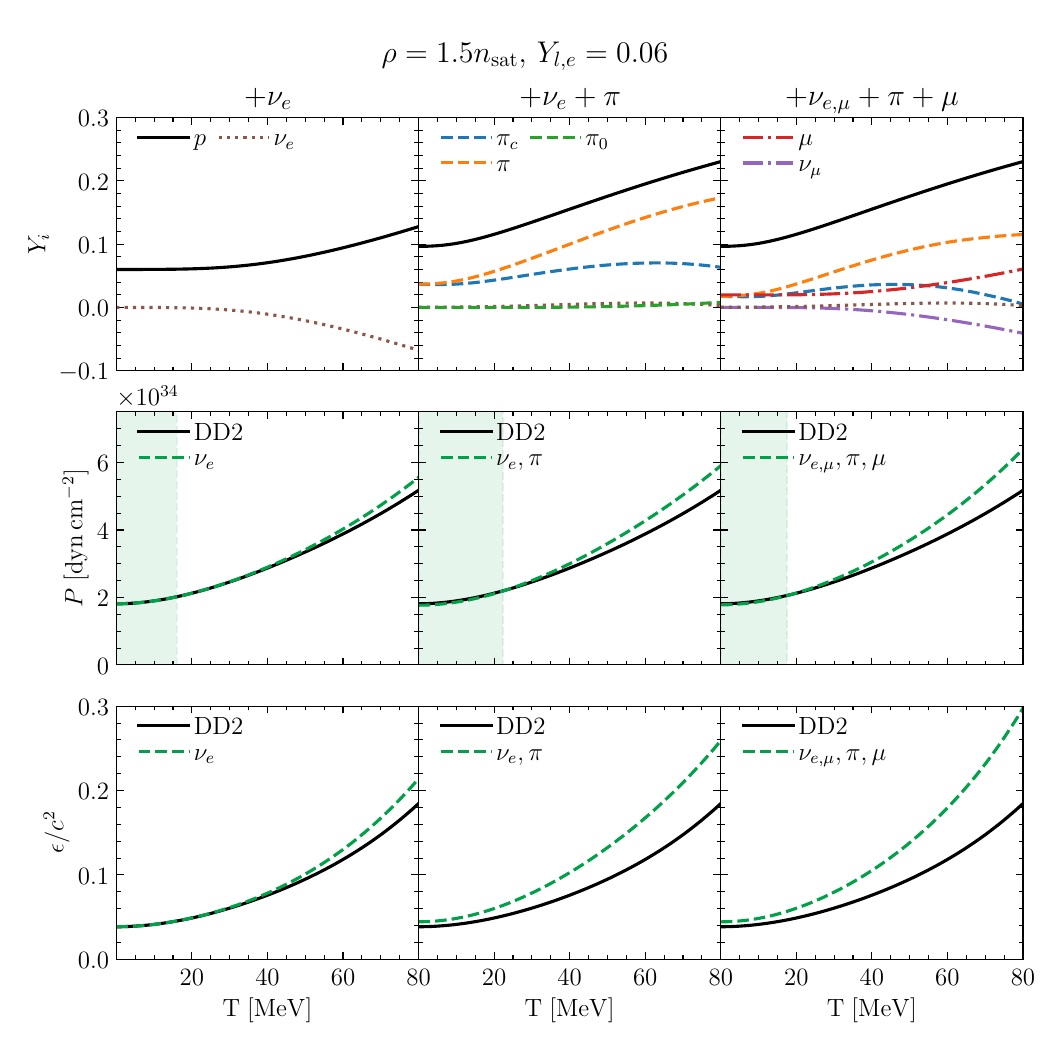}
    \caption{Same as Figure \ref{fig:9_panel_SFHo}, but for the DD2 EOS.}
    \label{fig:9_panel_DD2}
\end{figure*}

\subsection{General Equation of State Properties}
\label{sec:eos_properties}

In this Section, we present general properties of the $\nu_e-$, $\mu-$, and $\pi$-augmented equation of states, including particle fractions, pressure profiles, and specific internal energy profiles.
We begin our analysis to look at general properties of the EOS by investigating estimates for particle fractions $Y_i$ under typical thermodynamic conditions in the merger.  
These plots are generated for typical conditions found in a neutron star merger remnant \cite{Perego:2019adq,Hammond:2021vtv}, $\rho = 1.5 n_\mathrm{sat}$ = $3.9\times 10^{14}$ g cm$^{-3}$, $Y_{l,e} = 0.06$, and $0.01 \lesssim T \lesssim 80$ MeV.  These calculations are presented in Figure \ref{fig:9_panel_SFHo} and Figure \ref{fig:9_panel_DD2} for the SFHo and DD2 EOS, respectively.  

Begin by examining the top left panel of Figure \ref{fig:9_panel_SFHo}.  We see the particle fractions for protons $(Y_p)$ and electron type neutrinos $Y_{\nu_e}$.  Recall in this work, we define $Y_{\nu_e}$ (and similarly $Y_{\nu_\mu}$) as the difference of the particle fraction of the neutrinos with the particle fraction of the corresponding antineutrinos, Equation (\ref{eq:mu_lepton}).  This definition implies that negative values of $Y_{\nu_e}$ signify the presence of more antineutrinos than neutrinos.  At low temperatures, there is no noticeable imbalance between neutrinos and antineutrinos.  With increasing temperature, there is an increasing imbalance, indicated by $Y_{\nu_e} \sim -0.07$ at $T = 80$ MeV.  The $Y_p$ behavior is characterized by a steady increase with temperature from $Y_p = 0.06$ to $Y_p \sim 0.13$. This behavior is defined by Equation (\ref{eq:e_lepton}).  Since the table is defined at a fixed $Y_{l,e} = 0.06$, as $Y_{\nu_e}$ steadily decreases, $Y_e$, and thus $Y_p$, must steadily increase.

In the middle panel of the first row, we additionally examine  pion-related quantities for the $+\nu_e + \pi$ case.  The newly labeled quantities are condensed pions $(Y_{\pi_c})$, thermal and condensed charged pions ($Y_\pi$),  and thermal neutral pions $Y_{\pi_0}$.  At lower temperature values, the dashed orange line displays values around 0.04 for $Y_\pi$.  This is due to pion condensate forming.  As temperatures increase, the thermal population of pions increases.  By contrast, the condensate does not increase as rapidly.  Thus, $Y_{\pi_c} / Y_\pi$ is decreasing as temperature increases.  This behavior represents pions moving from the lowest available energy state into higher energy levels, due to increased temperature.  At $T = 80$ MeV, we see $Y_{\pi_c} \sim 0.08$ and $Y_\pi \sim 0.18$.  There is a moderate increase in $Y_{\pi_0}$ due to thermal effects. 
 Note the drastically different behavior of $Y_{\nu_e}$.  In contrast with the $+\nu_e$ case, $Y_{\nu_e}$ remains close to 0.  This is a direct consequence of pion condensation via Equation (\ref{eq:chem}).  Because $\mu_\pi$ is bounded by $m_\pi$, and the pions are in strong equilibrium with the baryons, $\mu_n - \mu_p$ is also limited by $m_\pi$. This in turn limits $\eta_{\nu_e}$, limiting the imbalance between electron neutrinos and antineutrinos. As a consequence of these low $Y_{\nu_e}$ values, notice the similar shape of $Y_p$ (solid black line) and $Y_\pi$.  These two curves are shifted by the $Y_e \sim Y_{l,e} \sim 0.06$, as a result of Equation (\ref{eq:charge}). 

In the right panel of the first row, we additionally examine muon-related quantities for the $+\nu_{e,\mu} + \pi + \mu$ case.  At low temperatures, in the dashed - dot red line, we see a $Y_\mu \sim 0.02$---and at temperatures near 80 MeV---climbing to $Y_\mu \sim 0.06$.  This behavior is justified as higher temperatures increase the total available amount of thermal energy to produce new particles.  At lower temperatures, there is no noticeable imbalance between muon type neutrinos and muon type antineutrinos.  However, at higher temperatures, we see $Y_{\nu_\mu} \sim -0.04$.  Similar to before, $Y_\mu$ and $Y_{\nu_\mu}$ mirroring each other is a direct consequence of Equation (\ref{eq:mu_lepton}).  In this regime, $Y_{l,\mu}$ remains constant, so the sum of both dashed-dot lines remains a constant as well.  Compared to the $+\nu_e + \pi$ case, at all temperatures, $Y_\pi$ (dashed line) has shifted to lower values because of the increased presence of charged $\mu$, limiting the total available charge for pion condensate to form.  Furthermore, notice $Y_{\pi_c}$ is clearly non monotonic near the upper end of the temperatures.  In this regime, the formation of $\bar{\nu}_\mu$ drives up the production $\mu$, in order to keep $Y_{l,\mu}$ constant.  With an increasing $Y_\mu$, the condensed population of pions cannot form as prolifically, forcing $Y_{\pi_c}$ to decrease more quickly.

One clear observation is that $Y_p$ displays similar behavior between the $+\nu_e + \pi$ and $+\nu_{e,\mu} + \pi + \mu$ cases.  This feature is a consequence of the formation of a pion condensate (see Appendix \ref{sec:pion_condensate} for a discussion about pion condensation).  In particular, $Y_p$ is determined for a given $\rho$, $T$, $Y_{l,e}$ when the difference in chemical potentials between neutrons and protons exceeds the charged pion mass, or $\mu_n - \mu_p \geq m_{\pi^-}$.  Physically, this expression implies, in the presence of condensate, that the charge fraction is limited by the mass of the pion.  How much condensate forms will be limited by the amounts of other charge carriers: muons and electrons, both of which are affected by the presence of neutrinos.

The left column in the middle row shows the pressure contributions in the $+\nu_e$ case.  At lower temperatures, the pressures are nearly identical because there is no significant production of $\nu_e$.  Only at higher temperatures, with increasing production of (anti)neutrinos, do more noticeable deviations occur, around the $10 \%$ level.

\begin{figure*}[t!]
\centering
\begin{subfigure}[t]{0.45\textwidth}
  \centering
  \includegraphics[width=\linewidth]{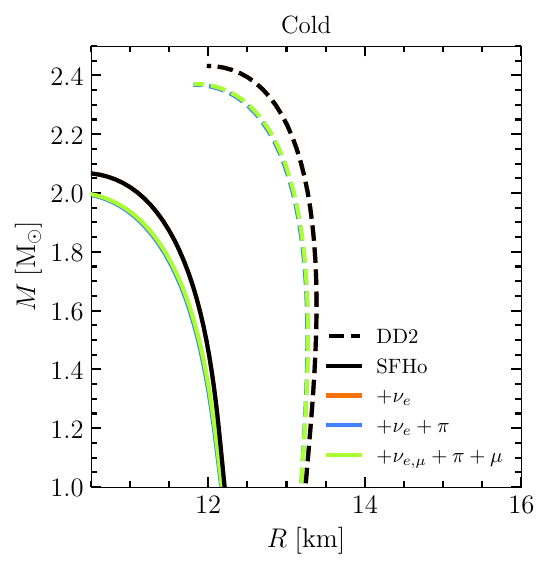}
  \label{fig:tov_cold}
\end{subfigure}%
\begin{subfigure}[t]{0.45\textwidth}
  \centering
  \includegraphics[width=\linewidth]{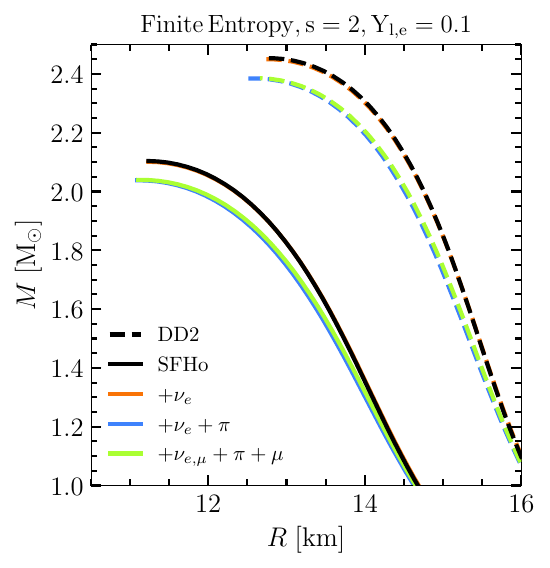}
  \label{fig:tov_both_s2}
\end{subfigure}

\caption{Mass ($M$)- radius ($R$) curves for non-rotating neutron stars.  Shown are results for the DD2 and SFHo equations of state.  The left column assumes cold $\beta-$equilibrium, and the right column assumes constant entropy of $s = 2 k_\mathrm{B} / \mathrm{baryon}$ at constant $Y_{l,e} = 0.1$. For all cases, the addition of pions ($\pi$) and muons ($\mu$), and trapped electron neutrinos ($\nu_e$) softens the EOS and creates smaller, less massive NSs.  
}
\label{fig:TOV}
\end{figure*}

The middle column, middle row shows pressures for $+\nu_e + \pi$.  At low temperatures, the pressure is slightly lower when pions are included, compared to the base EOS.  This is indicated by the green shaded region.  These results are in line with Ref. \cite{vijayan:2023}. This behavior is due to the fact that the increased presence of pions creates a larger proton fraction, compared to before.  At this density and $Y_{l,e}$, the pressure reaches a minimum at $Y_p \sim 0.08$.  Thus, the higher charge fraction near 0.1 is conducive to lower pressures.  Furthermore, the majority of pions at lower temperatures are condensate; they will not contribute to the pressure of the EOS.  At higher temperatures, the increased presence of thermal pions provides additional pressure, with the $+\nu_e +\pi$ case displaying higher pressures at $\gtrsim 5$ MeV.  In the middle row, right column, we notice the shaded region is removed.  This feature indicates the pressure is higher than the base EOS for the $+\nu_{e,\mu} + \pi + \mu$ case. While $Y_p(T)$ is similar for both EOS variants, the key difference is the increased presence of muons and $\nu_\mu$.  These particles are an additional source of pressure, causing slightly higher pressures at all temperatures, compared to the control EOS.  

The lower left plot displays the specific internal energy, normalized by the speed of light squared, as a function of temperature.  At low temperatures, we see the $+\nu_e$ case only marginally higher than the base EOS.  Similar to the pressure, only at higher temperatures does the increase in $\epsilon$ become apparent, reaching differences around the $20 \%$ level at the $T = 80$ MeV.

In the bottom row, middle column, we see $\epsilon$ behavior in contrast to \cite{vijayan:2023}, who see a lower energy density than the base EOS.  In our EOS table, at $\rho = 1.5 n_\mathrm{sat}$ and $T = 1$ MeV, we observe the behavior of $\epsilon$ as a function of $Y_p$.  We see a minimum in $\epsilon$ at $Y_p\sim 0.06$.  As $Y_p$ increases beyond that, so too does $\epsilon$.  This gives a $+\nu_e + \pi$ line higher than our control SFHo.  At higher temperatures, the $+\nu_e +\pi$ EOS continues to deviate from the base EOS as the energy contributions from the additional pion species create a larger $\epsilon$. In the last panel, the presence of muons increases the specific internal energies at higher temperatures through the existence of other species of muons and neutrinos.

Between both EOSs, both Figure \ref{fig:9_panel_SFHo} and Figure \ref{fig:9_panel_DD2} display similar qualitative behavior, with differences in the numerical values because of different stiffnesses of each EOS.  A noteworthy feature in Figure \ref{fig:9_panel_DD2} is the middle row describing the DD2 pressures.  The shaded regions, where the modified pressure is lower than the control pressure, are present for all three modifications of the EOS.  As new species are added, the temperature at which the modified pressure exceeds the control pressure varies from $\sim 16$ MeV to $\sim 22$ MeV to $\sim 17$ MeV for the $+\nu_e$, $+ \nu_e + \pi$, and $+\nu_{e,\mu} + \pi + \mu$ cases, respectively.  For $+\nu_e$ we do not expect this green shaded region, representing a pressure drop, to be dynamically relevant, because of the lack of $\nu_e$ in this low temperature regime.  The presence of pions in the middle panel produces a larger shaded region because of the formation of pressureless condensate.  The green shaded region in the right panel moves to lower temperatures because the additional muonic species provide additional pressure support, shifting the dashed green curve upwards.  In the lowest row, the behavior of $\epsilon$ behaves similar to SFHo.

\subsection{Influence on {Isolated} Neutron Stars}
\label{sec:tov}

To translate the changes in microphysics into macroscopic observables, we first construct cold, non-rotating neutron star solutions in weak-interaction ($\beta-$) equilibrium. To address the influence of neutrinos in a higher temperature scenario, we consider finite entropy, constant $Y_{l,e}$ NSs.

By solving the Tolman-Oppenheimer-Volkoff (TOV) equations \cite{tolman1939,oppenheimer:1939} for a variety of initial central densities, we construct mass-radius curves (Figure \ref{fig:TOV}). To construct these, we slice our finite temperature EOS for a variety of $\rho$ values, at a low temperature $T = 0.1$ MeV, imposing  $\beta-$equilibrium: $\mu_n - \mu_p = \mu_e$.  Note, we do not account for the neutrino chemical potential because trapped neutrinos are not prevalent at low temperatures.  These `cold' TOV stars are seen in the left panel of Figure \ref{fig:TOV}.

As expected for the cold case, including $\nu_e$ creates a mass-radius curve directly under the unmodified EOS results.  At this low temperature, there are not sufficient neutrinos to modify the NS structure, supported by the middle row, left panel of Figure \ref{fig:9_panel_SFHo}.  The inclusion of pions leads to an overall reduction in pressure, causing the EOS to soften considerably. As suggested in the middle row of Figure \ref{fig:9_panel_SFHo}, high density, low temperature environments will form pion condensate.  This translates to overall lower maximum masses in the EOS including pions, seen in the blue lines of Figure \ref{fig:TOV}.  Lastly, at marginally higher mass and radii, the $+\nu_{e,\mu} + \pi + \mu$ case (in lime green) behavior is justified by the additional pressure from muons.

Indeed, the post-merger remnant will have a region of high temperature in a ring structure around the center. To approximate relatively higher temperature in the interior of the star, compared to the outer atmosphere, we also construct beta equilibrium, finite entropy TOV stars, displayed in the right panel of Figure \ref{fig:TOV}.  To construct these, we begin with our EOS variants.  For a variety of densities, we select points that have entropies of $s = 2 \, k_\mathrm{B} / \mathrm{by}$ and have a constant $Y_{l,e} = 0.1$, coarsely representative of the remnant.
While muons do not appreciably change the mass-radius, the impact of trapped muon neutrinos, $\nu_\mu$, has been argued to affect the post-merger dynamics \cite{loffredo:2023}.
Trapped neutrinos only become relevant at high temperatures, as can be seen by the (small) changes in the mass radius curve when considering fixed specific entropy models, just below the black curves.  Similar to the cold TOV case, the presence of pions (specifically the condensate) has the largest impact on the structure of the mass-radius curves.



\subsection{Numerical Methods}
\label{sec:num_methods}

To process our EOS data, we use the `stellar collapse'\footnote{\url{stellarcollapse.org}} format, using a python analysis script.
Once our new EOSs are obtained, we solve the TOV equations using a publicly available solver \cite{steiner:2014}.

In this work, we also present results of neutron star merger simulations in full numerical relativity. To this end, we solve the equations of general-relativistic (magneto-)hydrodynamics (GRMHD) in dynamical spacetimes \cite{Duez:2005sf}.
We model the spacetime dynamics using the Z4c formulation of the Einstein equations \cite{Bernuzzi:2009ex,Hilditch:2012fp}. The combined set of equations is solved using the \texttt{Frankfurt/IllinoisGRMHD (FIL)} code \cite{Most:2019kfe,Etienne:2015cea}.  Alongside these equations, a leakage scheme is used to transport energy and lepton number in the optically thin regime \cite{Ruffert:1995fs,Galeazzi:2013mia}.  This scheme leads to changes in $Y_e$ and the energy of the matter in regions where neutrinos stream out of the remnant. \texttt{FIL} is based on the \texttt{EinsteinToolkit} infrastructure \cite{Loffler:2011ay}, and provides high-order conservative finite-difference methods for the GRMHD equations \cite{DelZanna:2007pk} and an unlimited fourth-order discretization for the spacetime equations \cite{Zlochower:2005bj}.
More specifically, \texttt{FIL} implements a fourth-order version of the ECHO scheme \cite{DelZanna:2007pk} using WENO-Z reconstruction \cite{borges2008improved}, and provides its own infrastructure for handling tabulated equations of state and primitive inversion using the method of \cite{Kastaun:2020uxr}. The computational domain is handled using the \texttt{Carpet} fixed mesh-refinement infrastructure \cite{Schnetter:2003rb}. In particular, we use $7$ levels of moving mesh refinement with a finest resolution $\Delta x = 260 \rm m$. The outer boundaries are placed at a distance of $3,000 \, \rm km$. The choice of resolutions is consistent with \texttt{FIL}'s beyond second-order convergence \cite{Most:2019kfe}, see also Refs. \cite{Most:2018eaw,Most:2019onn,Most:2021ktk,Most:2022yhe} for similar studies of nuclear matter effects with the code.

We generate numerical initial conditions using the \texttt{Kadath}\cite{Grandclement:2009ju}/\texttt{FUKA}\cite{Papenfort:2021hod} set of codes, which solve the extended conformal thin sandwich formulation \cite{Pfeiffer:2002iy}.
We neglect initial neutron star spins (see \cite{Tichy:2011gw}, and also Refs. \cite{Most:2020exl,Papenfort:2022ywx,Tootle:2021umi} for simulations studying the impact of spins with the code) and model the neutron stars as irrotational. We consider initial neutron star masses, $m_1 = 1.25 M_\odot$ and $m_2= 1.25 M_\odot$ at a separation of $45\, \rm km$.


\section{Simulation Results}
\label{sec:sim_results}

Following the construction of our augmented equation of state models, we now investigate their impact on fully general-relativistic neutron star merger simulations.


\begin{figure*}[]
    \centering
    \begin{subfigure}[t]{0.47\textwidth}
      \centering
      \includegraphics[trim={0cm 0cm 2.5cm 2cm}, clip, width=\linewidth]{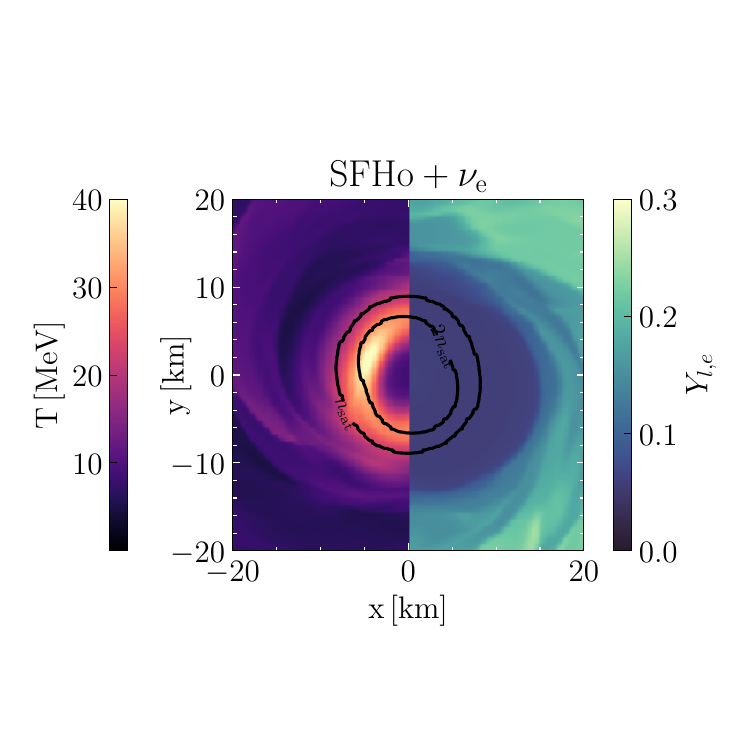}
          \end{subfigure}
    \begin{subfigure}[t]{0.47\textwidth}
      \centering
        \includegraphics[trim={2.5cm 0cm 0cm 2cm}, clip, width=\textwidth]{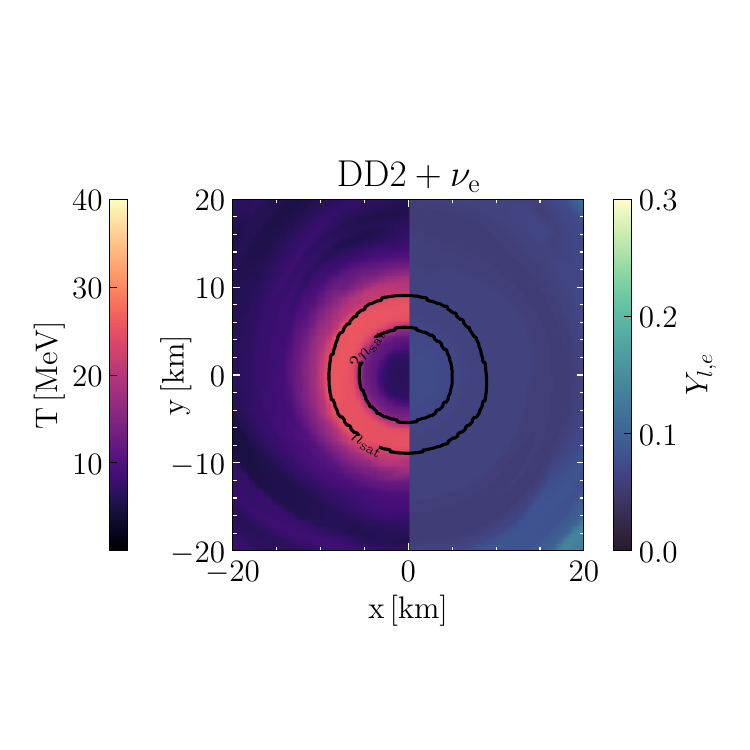}
    \end{subfigure}%
    \vspace{-2.3cm}

    \begin{subfigure}[t]{0.47\textwidth}
      \centering
      \includegraphics[trim={0cm 0cm 2.5cm 2.5cm}, clip, width=\linewidth]{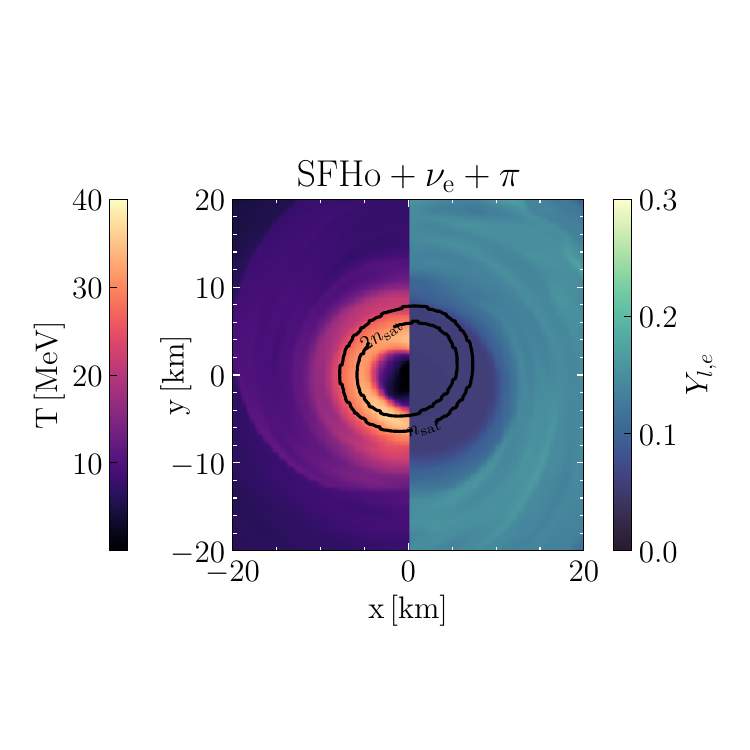}
          \end{subfigure}
    \begin{subfigure}[t]{0.47\textwidth}
      \centering
        \includegraphics[trim={2.5cm 0cm 0cm 2.5cm}, clip, width=\textwidth]{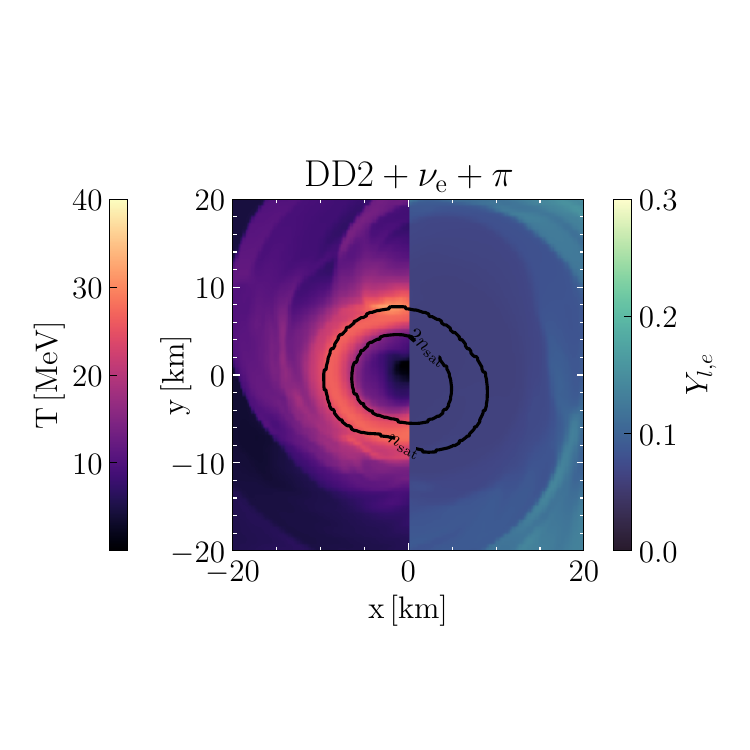}
    \end{subfigure}%
    \vspace{-2.3cm}

    \begin{subfigure}[t]{0.47\textwidth}
      \centering
      \includegraphics[trim={0cm 2cm 2.5cm 2.5cm}, clip, width=\linewidth]{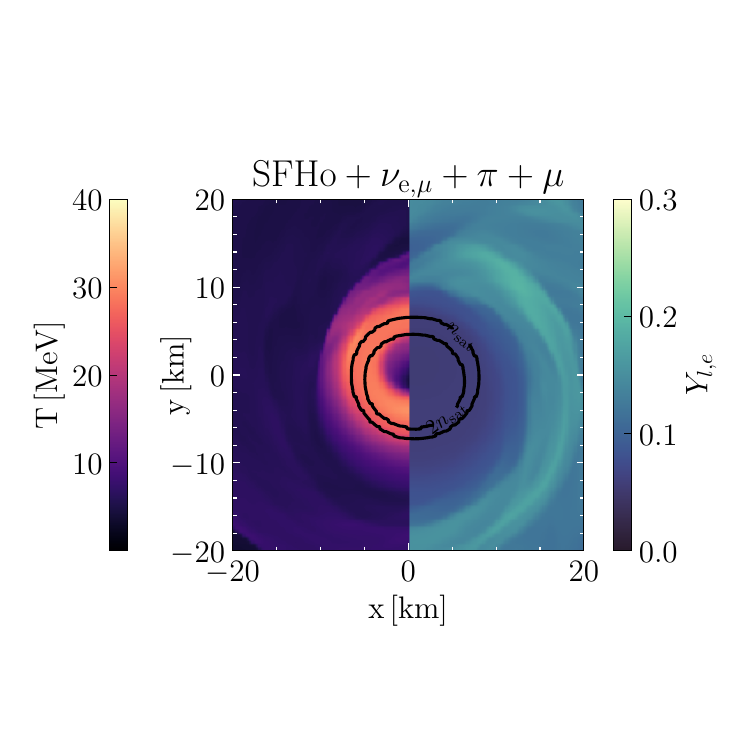}
          \end{subfigure}
    \begin{subfigure}[t]{0.47\textwidth}
      \centering
        \includegraphics[trim={2.5cm 2cm 0cm 2.5cm}, clip, width=\textwidth]{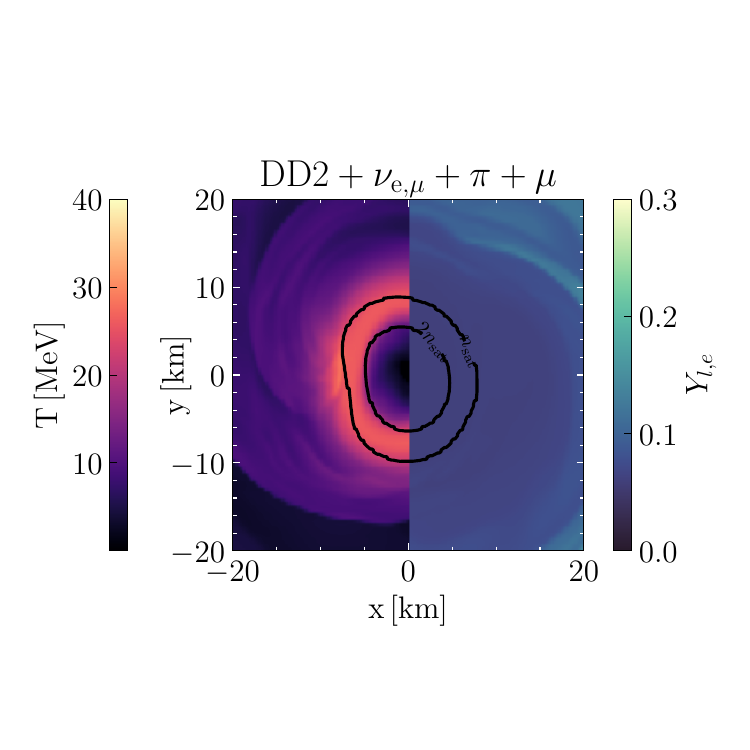}
    \end{subfigure}%

    \caption{Post-merger temperature, $T$, (left half) and electron lepton fractions, $Y_{l,e}$ (right half), taken $\sim 20$ ms after merger in the equatorial plane.  Each panel corresponds to an EOS variant.  Overlaid in black are the isodensity rest-mass contours at $n_\mathrm{sat}$ and $2 n_\mathrm{sat}$.}
    \label{fig:temperature}
    \end{figure*}

\begin{figure*}[t!]
\centering
\includegraphics[trim={0cm 1.5cm 0cm 0cm}, clip, width=0.9\textwidth]{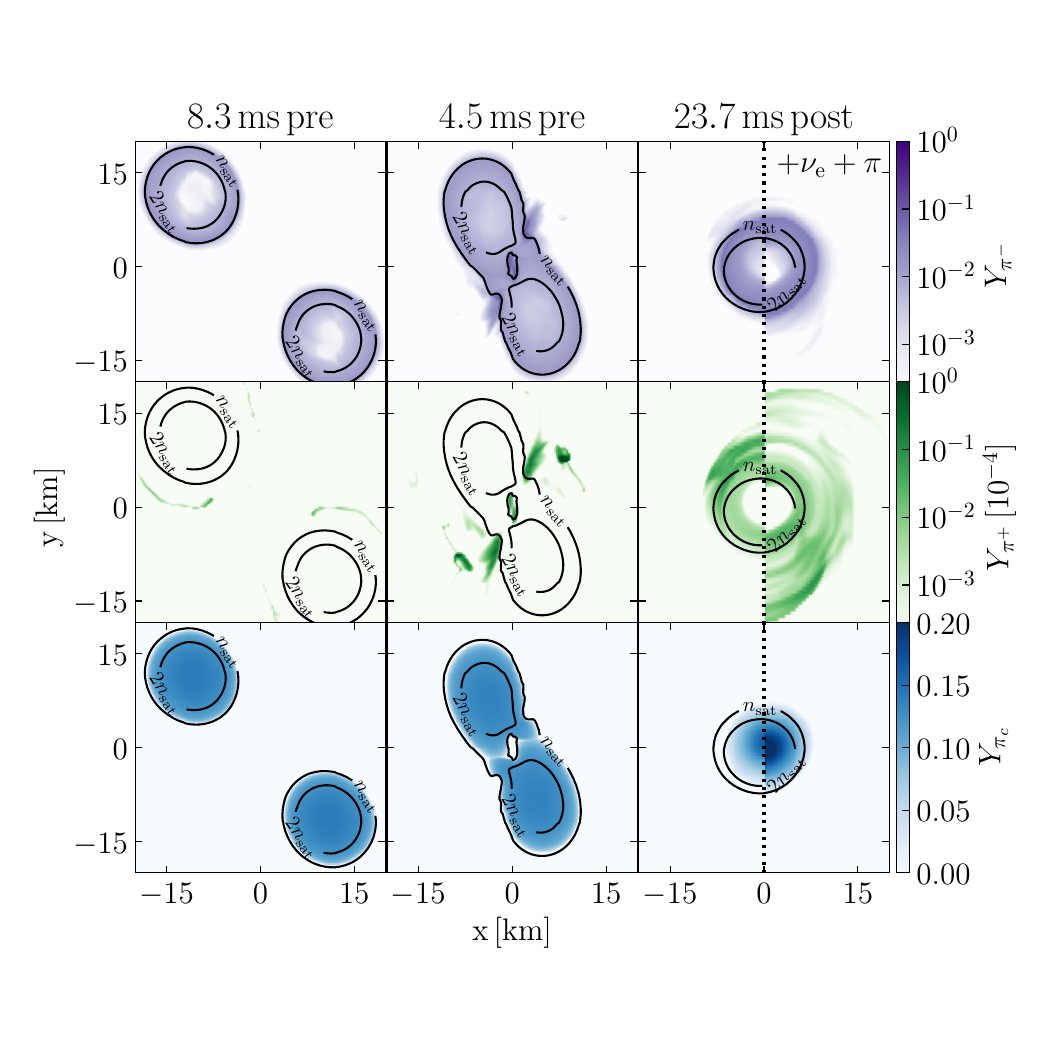}
\caption{Neutron star merger with trapped neutrinos, $\nu_{e,\mu}$, pions, $\pi$, and muons, $\mu$, using the SFHo equation of state. {\it (From left to right)} Equatorial slices of particle fractions, $Y_i$, before merger, at contact, and during ringdown.  (Top) Pion fraction of negatively charged thermal pions, $\pi^-$. (Middle) Pion fraction of positively charged thermal pions, $\pi^+$ (scaled by $10^4$).  (Bottom) Pion fraction for condensate, $\pi_c$.   Black solid lines are the rest mass density contours at $n_\mathrm{sat}$ and $2 n_\mathrm{sat}$.  In the last column, the left half of the panel includes muons whereas the right half of the column does not include muons.  Notice the lighter shade in the condensate indicates the presence of muons (left) lowers the condensate fraction.} 
\label{fig:slices_pions}
\end{figure*}

\begin{figure*}[t!]
\centering
\includegraphics[trim={0cm 1.5cm 0cm 0cm}, clip, width=0.9\textwidth]{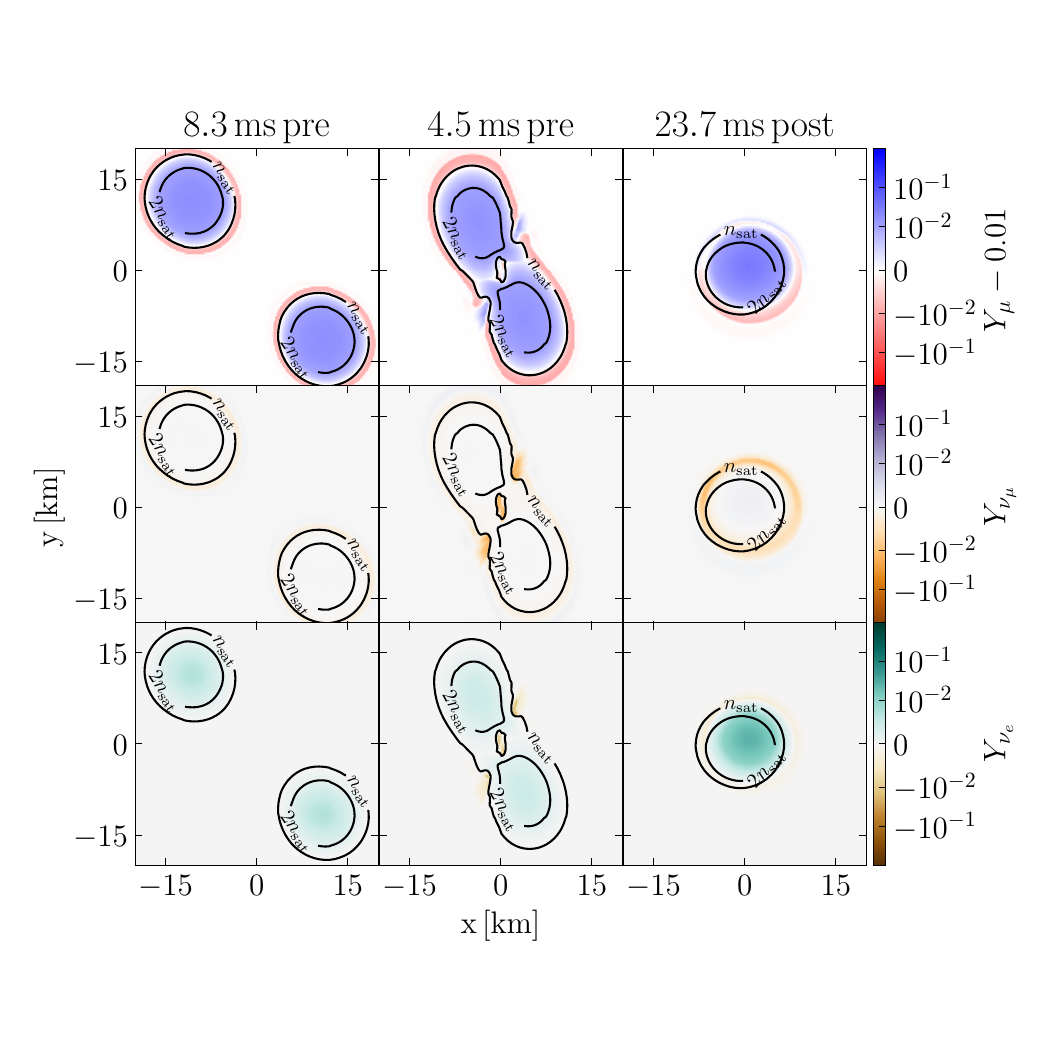}

\caption{Same as Figure \ref{fig:slices_pions} but for muon fraction, $Y_\mu$, muon neutrino fraction $Y_{\nu_\mu}$, and electron neutrino fraction $Y_{\nu_e}$. (Top) Muons are present in the cold isolated NSs and display an increased population in the remnant core as densities and temperatures increase. (Middle) Muon-type neutrino fraction.  As the isolated NSs are constructed assuming zero temperature equilibrium, very few muon-type neutrinos are present. 
 As temperatures increase at contact, an excess of $\bar{\nu}_\mu$ (negative $Y_{\nu_\mu}$) is apparent.  (Bottom) Electron-type neutrino fraction, displaying high $\nu_e$ in the remnant core and higher $\bar{\nu}_e$ content in the high temperature ring.} 
\label{fig:slices_muons}
\end{figure*}

\begin{figure}
    \centering
    \includegraphics[width=0.4\textwidth]{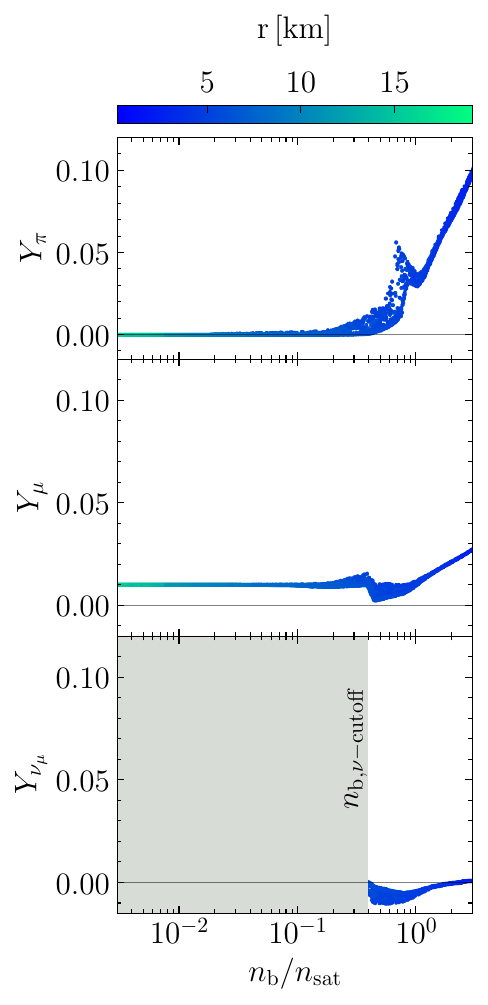}
    \caption{Particle fractions, $Y_i$, as a function of normalized number density, $n_b$, in merger remnant for model SFHo$+\pi+\mu + \nu_e$ $\sim 20$ ms after merger. Colors indicate the radial distance, $r$, from the center.
    }
    \label{fig:swan}
\end{figure}

\begin{figure}[t!]
    \centering
    \begin{subfigure}[t]{0.4\textwidth}
      \centering
      \includegraphics[width=\linewidth]{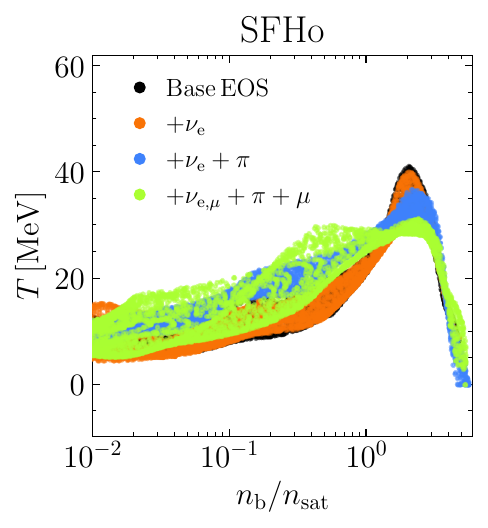}
          \end{subfigure}
    \begin{subfigure}[t]{0.4\textwidth}
      \centering
        \includegraphics[width=\textwidth]{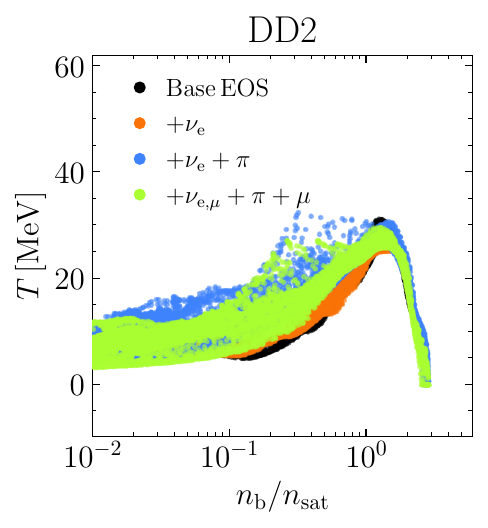}
    \end{subfigure}%
    \caption{Temperature, $T$, vs baryon number density, $n_b$, normalized by nuclear saturation density, $n_{\rm sat}$, for the post merger remnant for two different base equations of state (DD2 and SFHo), modified by the particle contributions listed in the legend. The data is taken at $\rm \sim 20 \, ms$ after merger.}
    \label{fig:temp_vs_nsat}
    \end{figure}

\begin{figure*}[t!]
    \centering
    \begin{subfigure}[t]{0.55\textwidth}
      \centering
      \includegraphics[trim={0 1.5cm 0cm 1.5cm}, clip, width=\linewidth]{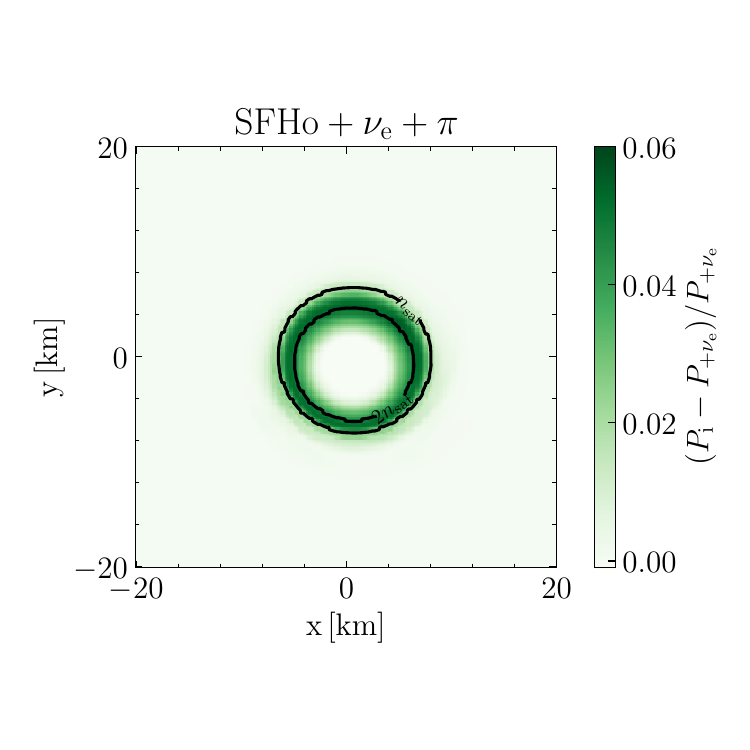}
    \end{subfigure}
    \hspace{-2.5cm}
    \begin{subfigure}[t]{0.55\textwidth}
      \centering
        \includegraphics[trim={0 1.5cm 0cm 1.5cm}, clip, width=\textwidth]{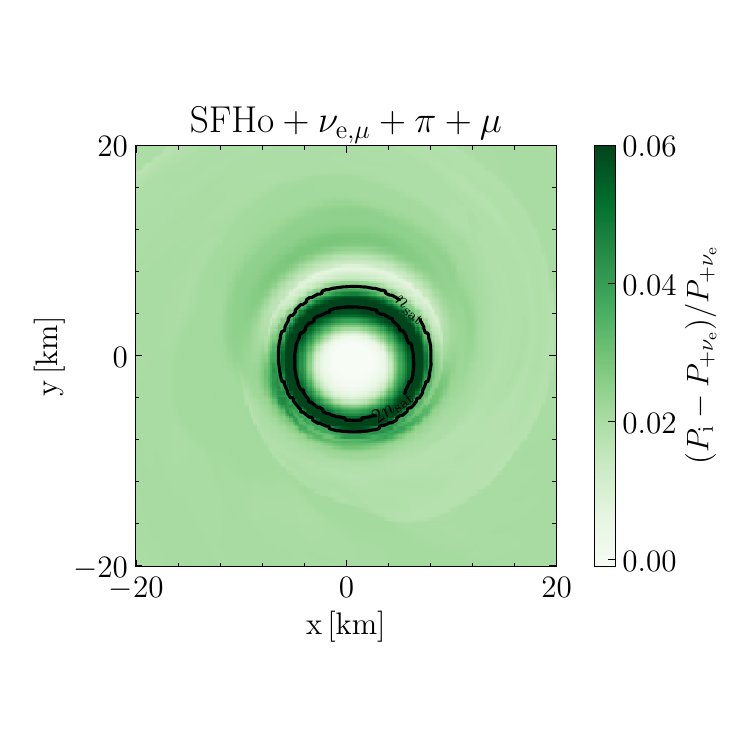}
    \end{subfigure}    

    \begin{subfigure}[t]{0.55\textwidth}
      \centering
      \includegraphics[trim={0 1.5cm 0cm 1.5cm}, clip, width=\linewidth]{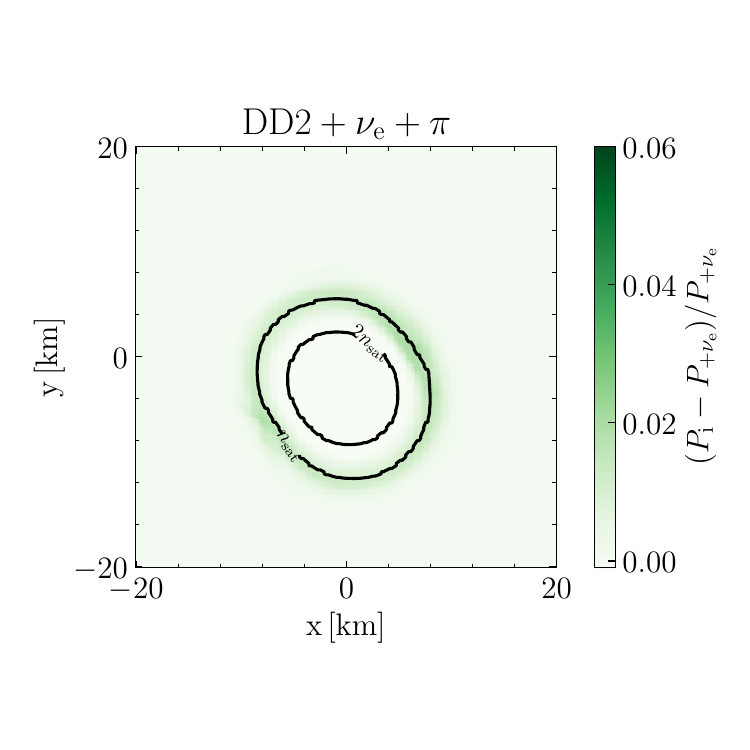}
    \end{subfigure}
    \hspace{-2.5cm}
    \begin{subfigure}[t]{0.55\textwidth}
      \centering
        \includegraphics[trim={0 1.5cm 0cm 1.5cm}, clip, width=\textwidth]{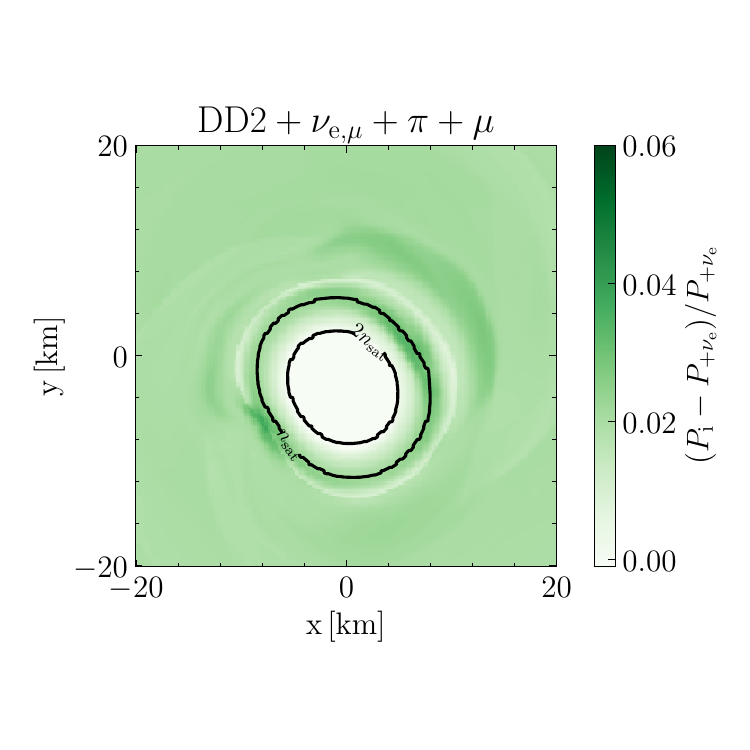}
    \end{subfigure}  
    
    \caption{Relative pressure difference as a result of different particle species. $P_j$ denotes the pressure given a composition. Shown are results for two EOS (DD2 and SFHo) about $\sim 20$ ms after merger when comparing $+\nu_e + \pi$ vs $+\nu_e$ (Left) and $+\nu_{e,\mu}+\pi+\mu$ vs $+\nu_e$ (Right). Overplotted with black lines are the density contours for $n_\mathrm{sat}$ and $2 n_\mathrm{sat}$. }
    \label{fig:prim_slice}
    \end{figure*}


\begin{figure*}
    \centering
    \begin{subfigure}[t]{\textwidth}
      \centering
      \includegraphics[width=\linewidth]{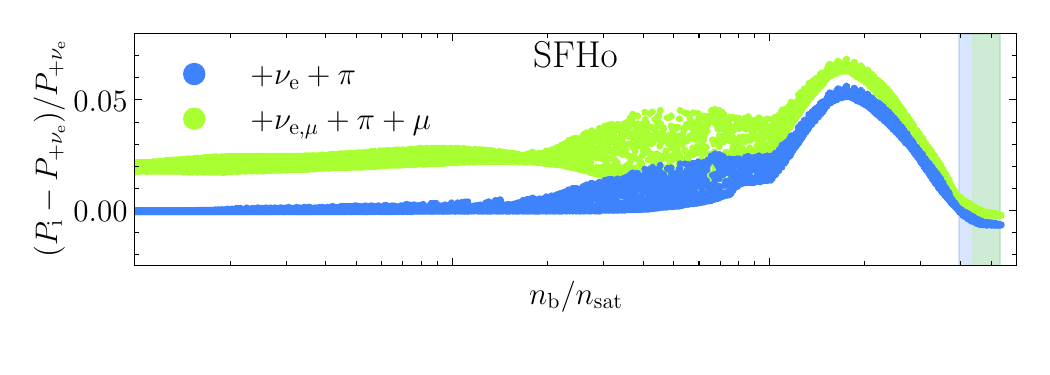}
          \end{subfigure}
    \vspace{-2.2cm}

    \begin{subfigure}[t]{\textwidth}
      \centering
        \includegraphics[width=\textwidth]{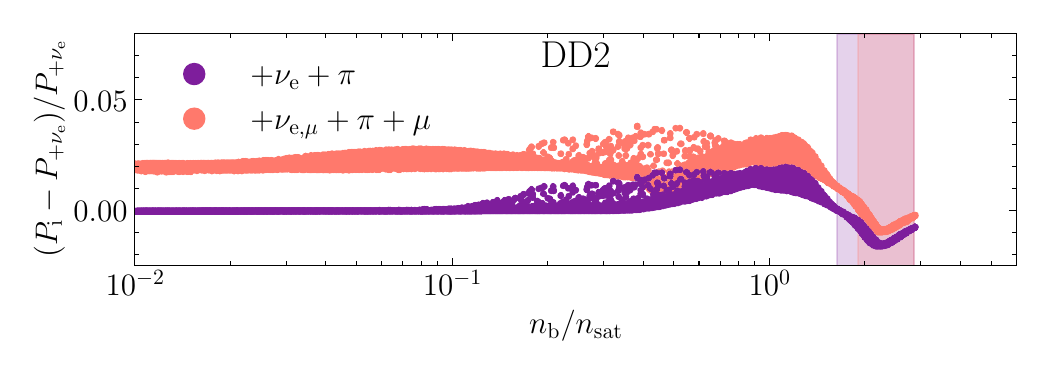}
    \end{subfigure}%
    \caption{Change in pressure in remnant for different SFHo EOSs (top) $\sim 20$ ms post merger and the DD2 EOSs (bottom).  Each color represents the difference in pressure for the $+\nu_{e,\mu} + \pi + \mu$ and $+\nu_e + \pi$ compared against $+\nu_e$. Shaded regions indicate where pressures for the labeled EOS are lower than the $+\nu_e$ case.}
    \label{fig:delta_p_vs_rho}
\end{figure*}



\subsection{Merger Dynamics and Remnant Composition}
\label{sec:composition}

Having identified how the initial conditions of the individual neutron stars will change, we transition to describing where in the merger and post-merger these species should occur and how this region changes throughout the evolution of the system.  To provide a picture of the merger remnant, we focus on Figure \ref{fig:temperature}. 
 The right half of each panel shows the electronic lepton fraction, $Y_{l,e}$.  Owing to initial weak-interaction equilibrium, the system commonly features low $Y_{l,e}$ cores and relatively higher $Y_{l,e}$ outer disk material. 
 For all cases, we see a high temperature ring up to $40$($30$) MeV off-center for SFHo (DD2). We can see that the addition of $\pi$ leads to a small reduction in overall temperatures. Likewise, we provide density contours at $n_\mathrm{sat}$ and $2 n_\mathrm{sat}$, seen in black.  This reveals that the SFHo merger with $\pi$ is slightly more compact. This remnant picture will be useful in describing the particle fractions for the new species.   Differences in the structure of the temperature profiles will be discussed in Section \ref{sec:dynamics}.

 Figure \ref{fig:slices_pions} describes the particle fractions of the negatively charged thermal pions, positively charged thermal pions, and pion condensate.  Due to the low temperature in the inspiral, the initial NSs prior to merger feature only small amounts of (thermal) pions, $Y_{\pi^-} \lesssim 10^{-2}$. The merger itself generates high temperatures via shock heating (Fig. \ref{fig:temperature}), leading to a dynamical production of pions (see also \cite{vijayan:2023}). Positively charged pions, $\pi^+$, are largely absent in the inspiral, 
 but are produced in small amounts in the post-merger, $Y_{\pi^+} \sim 3\times 10^{-6}$, largely in regions of high temperature or at lower densities in the disk.
 On the other hand, most pions especially in very dense regions $n\gg 2 \, n_{\rm sat}$, form a condensate, $\pi_c$, after merger.
The presence of muons, $\mu$, denoted by the split panel in Fig. \ref{fig:slices_pions}, leads to a reduction in the pion condensate, consistent with muons being produced by pion-decay. This can be seen by overall local charge conservation, where the presence of muons leads to a necessary reduction in pions, as the baryon number (mainly constraining the protons) is always fixed.

We now analyze the muons, muon-type (anti)neutrinos, and electron type (anti)neutrinos (Figure \ref{fig:slices_muons}). We can see that the muon fraction initially does not appreciably change from the initial value, $Y_\mu=0.01$, we had fixed to a representative value \cite{loffredo:2023}.  During and after merger, $Y_\mu$ increases to between $Y_\mu \sim 0.027-0.045$ from higher muon chemical potentials in this hotter and denser environment. Note, this is similar to the results from \cite{gieg:2024}, whose `5 neutrino species' cases find significant muons only inside the high density remnant core.


In the middle row of Figure \ref{fig:slices_muons}, we show the difference between $Y_{\nu_\mu}$ and $Y_{\bar{\nu}_\mu}$.  In the left panel, between both density contours, one sees a slight preference towards $\bar{\nu}_\mu$.  With maximum values of $|Y_{\nu_\mu}| \lesssim 3\times 10^{-3}$, these effects are minor enough to be dynamically unimportant.  The surface of the stars features small amounts of $\bar{\nu}_\mu$, which may, however, be contaminated by artificial heating at the surface, due to our numerical scheme not being well-balanced---unable to maintain hydrostatic equilibrium in the stellar atmosphere to machine precision \citep{kappeli:2016,gaburro:2021}.  In the middle row, middle column, we observe the production of $Y_{\nu_\mu}$ at the hot merger interface.  The middle row, right column displays a much more prominent `anti neutrino ring'.  This feature very closely tracks the high temperature ring in the remnant. Since the center of the remnant is relatively cooler (see Figure \ref{fig:temperature}), the production of $\bar{\nu}_\mu$ is less prominent.

In the lower row of Figure \ref{fig:slices_muons}, we show the difference between $Y_{\nu_e}$ and $Y_{\bar{\nu}_e}$.  The nearly isolated neutron stars show negligible $\nu_e$ contributions.  At contact, there is slight electron type neutrino production along the linear, diagonal, high temperature element.  In the right column, we see higher $\nu_e$ concentrations in the center of the remnant.   Note, for all three $\nu_e$ plots, the $\nu_e$ fractions track the $\mu^-$ fractions rather closely.  This is a product of the muons decaying into $\bar{\nu}_e$ in Equation (\ref{eq:mu_decay}). 
Overall, our results are consistent with previous works \cite{loffredo:2023,perego:2019}.


To better quantify the particle fractions in the remnant, at $\sim 20$ ms after merger, we refer to Figure \ref{fig:swan}.  Focus on the top panel of Figure \ref{fig:swan}.  This displays $Y_\pi$ as a function of number density $n_\mathrm{b}$, normalized by $n_\mathrm{sat}$, for the SFHo$+\nu_{e,\mu} + \pi + \mu$ EOS. 
It is clear in the high density regime the presence of charged pions is greater, supporting our previous figures.  Moreover, there is an inherent temperature-induced spread in the relation between $n_\mathrm{b}$ and $Y_\pi$.  In particular, the shocked, higher temperature regime near the center of the remnant will exhibit multiple $Y_\pi$ for a given $n_\mathrm{b}$. 

 Instead of a monotonic increase of $Y_\pi$ with $n_\mathrm{b}$, there is nonmonotonicity $\sim 0.75 n_\mathrm{sat}$.  Recall that the contributions from $Y_\pi$ come from both condensate and thermal effects.  When observing the plots for these species individually, we note a monotonic increase of condensate fraction with $n_\mathrm{b}$.  However, the thermal, charged pions exhibit the nonmonotonicity. This can be explained from the temperature profile within the remnant.  As outlined in our methods, once a condensate begins to form (around $n_\mathrm{b} / n_\mathrm{sat} = 1$) the chemical potential of the thermally charged pions remains fixed at $m_\pi$.  Referring to Equation (\ref{eq:n_bose}), this implies $z = 1$.  Thus, $n_\pi$ will scale as $T^{3/2}$.  Because of the presence of a high temperature ring around the remnant center (Figure \ref{fig:temperature}), as one approaches the center, there will be nonmonotonic behavior in the temperature as well.

In the middle panel of Figure \ref{fig:swan}, $Y_\mu$ as a function of $n_\mathrm{b}$ is shown.  For low densities, the constant $Y_\mu = 0.01$ is displayed.  However, at the density cutoff of $10^{14}$ g cm $^{-3}$, there is a decrease, followed by an increase at $\sim 0.45 n_b$.  The origin of this nonmonotonicity is different than the pion case.  As detailed in Appendix \ref{sec:ylmu_cold}, at this density, we transition from the assumed $Y_\mu = 0.01$ to the parameterized $\tilde{Y}_{l,\mu}(\rho)$ in Equation (\ref{eq:ymu_cold}). Just above the density cutoff, $\tilde{Y}_{l,\mu}(\rho)$ admits $Y_\mu < 0.01$, resulting in the first decrease.  At $n_\mathrm{b} \sim 0.45 n_\mathrm{sat}$, $\tilde{Y}_{l,\mu}(\rho)$ allows higher values, resulting in a corresponding increase.  

The lower panel displays $Y_{\nu_\mu}$.  Once one passes the density cutoff of $\rho \simeq 10^{14}$ g cm$^{-3}$, there is a minimum in $Y_{\nu_\mu}$ occurring at a similar density regime to the nonmonotonicity in $Y_\pi$.  Recall Equation (\ref{eq:nu_number}).  Note, this scales with $T^3 (F_2(\eta_\nu) - F_2(-\eta_\nu))$.  At these densities, the temperature inversion 
 is sufficient to prevent the $Y_{\nu_\mu}$ from becoming more negative.  When looking at the DD2 data, similar features in the particle fractions are present.

\subsection{{Quantifying} Merger Dynamics}
\label{sec:dynamics}


To begin describing the dynamics of the merger, we appeal to the temperature distribution as a function of $n_b / n_\mathrm{sat}$ in Figure \ref{fig:temp_vs_nsat}.  For all cases, we see increasing temperature with increasing number density of the baryons, up until $\sim 2 n_\mathrm{sat}$.  Beyond this density cut, the temperature decreases.  This observation is consistent with the high temperature ring between $n_\mathrm{sat}$ and $2 n_\mathrm{sat}$, seen in Figure \ref{fig:temperature}.  For both the base EOS and $+\nu_e$ case, we see no major changes in the thermal distribution of the material, indicating $\nu_e$ has no major effect on the thermalization of the new remnant at low densities, when approximated as an ideal Fermi gas.

A more subtle feature occurs at the peak of the temperature distribution; we see a marginally smaller peak temperature in the $+\nu_e$ case.  At contact between the no-longer isolated NS companions, there is a temperature spike that is of order 50 MeV.  Combined with the high density material, this high temperature environment leads to a trapped (anti) neutrino population.  Thus, the neutrinos provide a thermal energy sink for the merger. 

\begin{figure*}
    \centering
    \includegraphics[width=\linewidth]{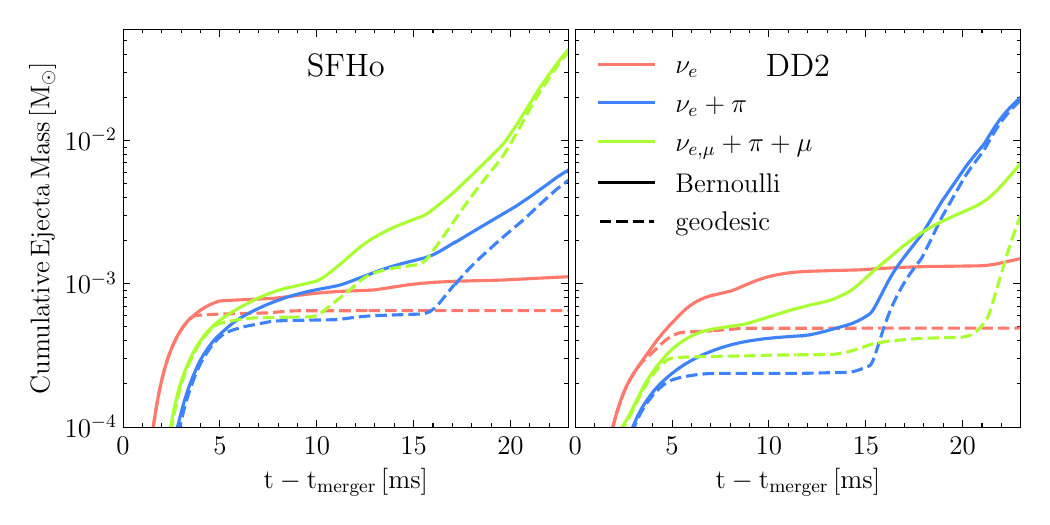}
    \caption{Time evolution of cumulative mass ejecta for SFHo (left) and DD2 (right).  Solid lines correspond to calculations according to Bernoulli's criteria  and dashed lines use the geodesic criteria for unbound mass.  The base EOS produces evolution similar to the $\nu_e$ case.}
    \label{fig:mass_ejecta_vs_t}
\end{figure*}

\begin{figure*}[t!]
    \centering
      \centering
    \begin{subfigure}[t]{\textwidth}
      \centering
      \includegraphics[width=\linewidth]{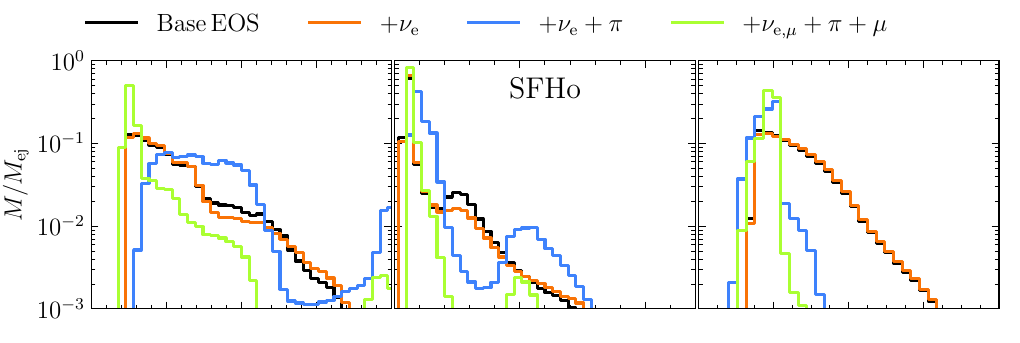}
          \end{subfigure}
    \vspace{-0.5cm}

    \begin{subfigure}[t]{\textwidth}
      \centering
        \includegraphics[width=\linewidth]{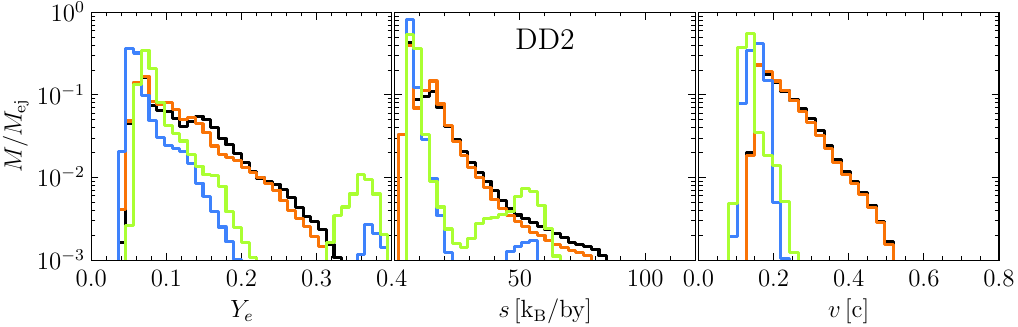}
    \end{subfigure}%
    \caption{Ejected mass fraction $M$ relative to the total ejected mass $M_{\rm ej}$ for electron lepton number ($Y_{e}$), entropy per baryon ($s$), and velocity ($v$).}
    \label{fig:mejection}
\end{figure*}

For the $+\nu_e + \pi$ case, we see a similar increase in temperature as a function of $n_b$.  Between $10^{-2}\, n_\mathrm{sat}$ and $n_\mathrm{sat}$, we see slightly higher temperature material at the $\sim 5\%$ level.  Beyond $\sim 2 n_\mathrm{sat}$ the high temperature peak is lower, from $\sim 40$ MeV to $\sim 35$ MeV, compared to the $+\nu_e$ case.  This lower peak is due to the thermal production of pions.  This observation is supported by the top panel of Figure \ref{fig:swan}.  Near $n_\mathrm{sat}$, there is a local maximum that displays a nonmonotonicity in $Y_\pi$.  This is directly sourced by the thermally charged pions, whereas the overall monotonic increase is sourced by the condensed pions.  With a relatively greater population of new species in a high temperature regime, the energy can go into creating $\pi$ rather than increasing the temperature of the baryons.

The $+ \nu_{e,\mu} + \pi + \mu$ case shows similar behavior, with slightly larger spread in temperatures between $10^{-2} n_\mathrm{sat}$ and $n_\mathrm{sat}$.  This may be due to the increased presence of muons.  In particular, having an additional species present may allow for further particle production, acting as an additional sink for the thermal energy.  

Focusing on the bottom panel of Figure \ref{fig:temp_vs_nsat} for DD2, we see some qualitative similarities to SFHo.  First, there is an increase in temperature vs $n_\mathrm{b}$ until $\sim n_\mathrm{sat}$.  Secondly, we do not see a major difference between unmodified DD2 and the $\nu_e$ case.  Third, we see thermalization of the lower density material due to pions, albeit at a more extreme extent, compared to SFHo.  This is illustrated by a more noticeable difference between the blue $+\nu_e + \pi$ points and the black DD2 points. This is counterbalanced with the inclusion of muons, leading to overall lower temperatures.

To better quantify what is driving the changes in the dynamics, we now analyze the contribution of different particle specifies via their pressure contributions.
To this end, we perform a similar analysis previously done for the impact of neutrino-driven effects in the merger \cite{Most:2021zvc,Most:2022yhe}. Defining $P_i$ as the pressure for an EOS with additional $i-$species, we compute $\Pi/P_{+\nu_e} = \left(P_i - P_{+\nu_e}\right)/P_{+\nu_e}$ as the relative pressure difference to our fiducial model including trapped electron neutrinos \cite{Espino:2023dei}. The outcome is shown in Fig. \ref{fig:prim_slice}.  This figure is taken $\sim 20$ ms after merger for the $+\nu_{e,\mu}+\pi+\mu$ run for SFHo (top) and DD2 (bottom).  It compares the pressures using the thermodynamic states found using the $+\nu_e$ run, with the $+\nu_e + \pi$, and $\nu_{e,\mu} + \pi + \mu$ cases.  
   We note a max $\Pi/P_{+\nu_e}$ in the remnant in a ring $\sim 7$ km from the remnant center at the $\sim 6 \%$ level.  The left column compares $+\nu_e$ with $+\pi+\nu_e$---essentially this comparison isolates the effects of the pions.  While we see this region fitting nicely in between the $n_\mathrm{sat}$ and $2 n_\mathrm{sat}$ contours, it is in fact the temperature that is sourcing the pressure change.  Figure \ref{fig:temperature} displays the temperature of the remnant.  Physically, the temperature is sourcing the thermal population of pions, which provides support to the remnant.   We see nice agreement between the temperature and pressure increases between Figure \ref{fig:temperature} and all panels of Figure \ref{fig:prim_slice}. 

\begin{figure*}[]
    \centering
    \begin{subfigure}[t]{0.47\textwidth}
      \centering
      \includegraphics[trim={0cm 0cm 2.369cm 0cm}, clip, width=\linewidth]{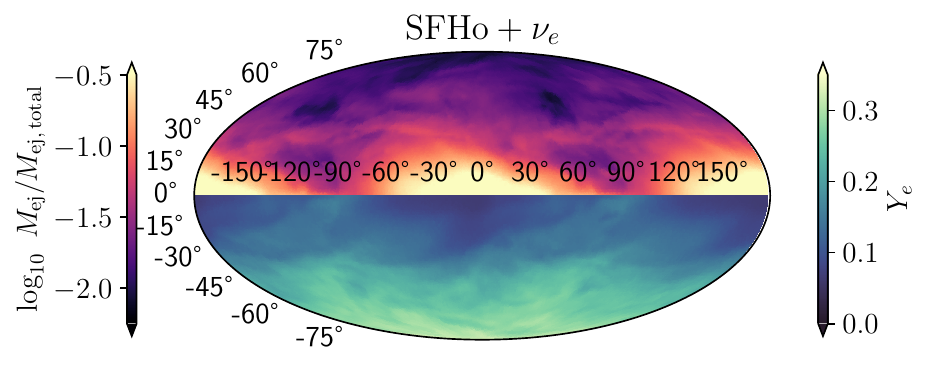}
          \end{subfigure}
    \begin{subfigure}[t]{0.47\textwidth}
      \centering
        \includegraphics[trim={2.369cm 0cm 0cm 0cm}, clip, width=\linewidth]{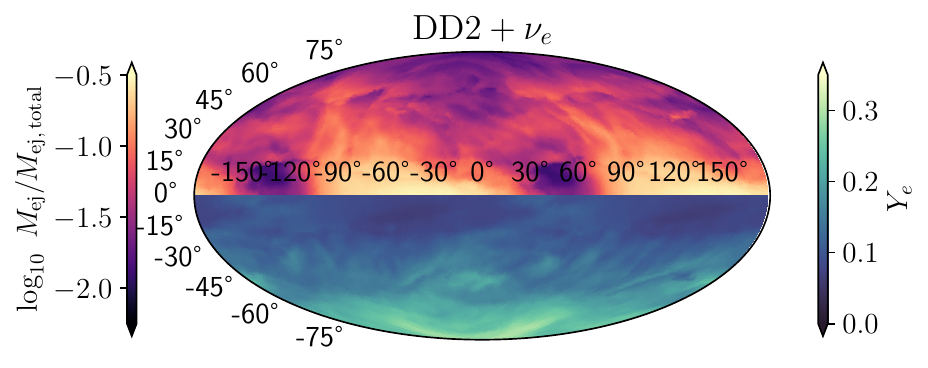}
    \end{subfigure}%

    \begin{subfigure}[t]{0.47\textwidth}
      \centering
      \includegraphics[trim={0cm 0cm 2.369cm 0cm}, clip, width=\linewidth]{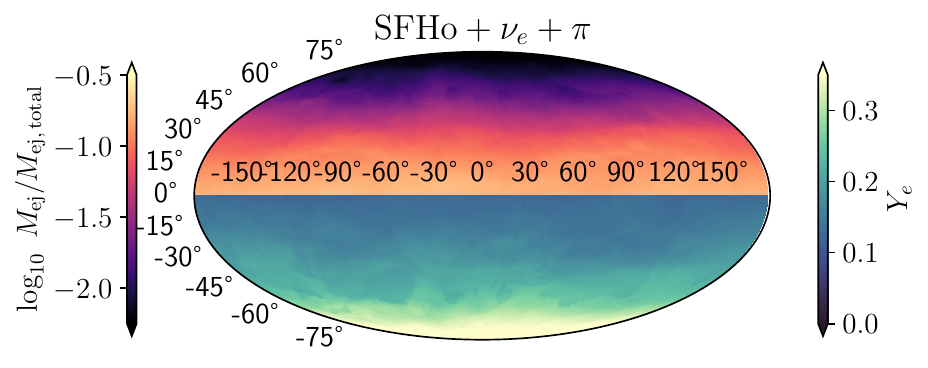}
          \end{subfigure}
    \begin{subfigure}[t]{0.47\textwidth}
      \centering
        \includegraphics[trim={2.369cm 0cm 0cm 0cm}, clip, width=\linewidth]{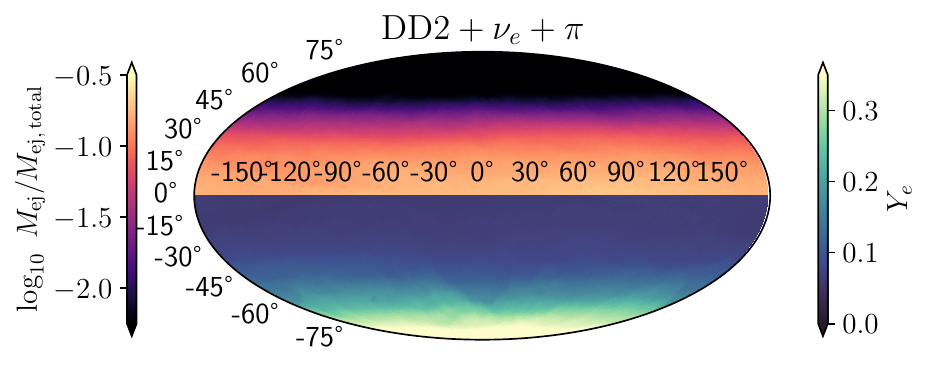}
    \end{subfigure}%

    \begin{subfigure}[t]{0.47\textwidth}
      \centering
      \includegraphics[trim={0cm 0cm 2.369cm 0cm}, clip, width=\linewidth]{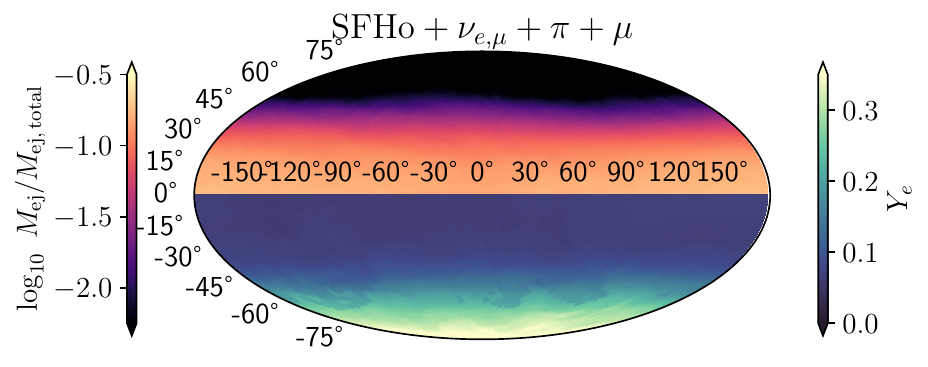}
          \end{subfigure}
    \begin{subfigure}[t]{0.47\textwidth}
      \centering
        \includegraphics[trim={2.369cm 0cm 0cm 0cm}, clip, width=\linewidth]{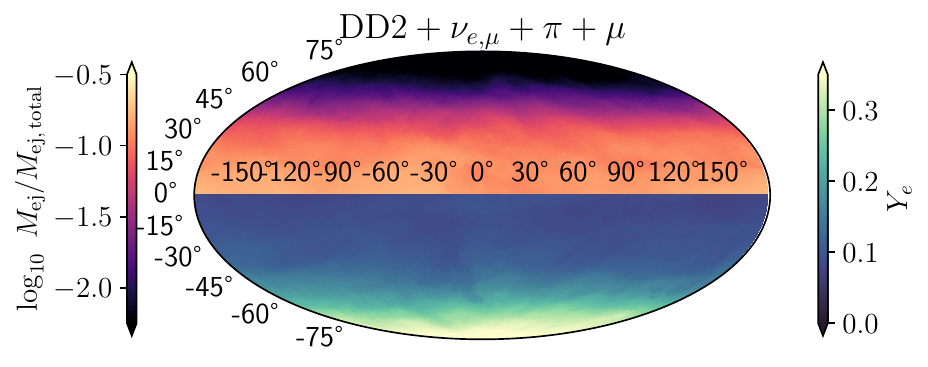}
    \end{subfigure}%

    \caption{In each panel, ejecta mass $M_\mathrm{ej}$(top) and $Y_{e}$ distribution (bottom) for all 3 EOS variants.  Brighter colors indicate more mass ejected and higher $Y_{e}$ along a particular direction.  Generally, more matter, which is neutron rich (low $Y_{e}$) is ejected along the equator, leaving darker $Y_{e}$ colors along the equator. }
    \label{fig:ejecta_spatial_distribution}
    \end{figure*}
 
In the right column of Figure \ref{fig:prim_slice}, we compare $+\nu_{e,\mu}+\pi+\mu$ with the $+\nu_e$ case.  This comparison isolates the effect of the pions and muons.  We see similar behavior to the previous case, with pressure increases at the $\sim 6 \%$ level in the high temperature ring surrounding the remnant center, still influenced by the thermal pions.  However, when comparing the left column with the right column, one notices the effect of the muons and muon type (anti)neutrinos.  Strikingly, there is a darker shade in the surrounding remnant material outside the contours at the $\sim 2\%$ level.  This is a direct consequence of the trends seen in the middle row of Figure \ref{fig:swan}.  The nonzero population of muons at lower density provide thermal support in the lower density material.  When comparing the SFHo cases in the top row with the DD2 cases in the bottom row, we can see that the impact is more pronounced for the SFHo EOS, due to the higher temperatures probed in the merger.

To better quantify these relations, Figure \ref{fig:delta_p_vs_rho} displays $\Pi/P_{+\nu_e}$ as a function of $n_\mathrm{b}$.  We begin by examining the blue $+\nu_e + \pi$ points; these correspond to the data in the upper left panel of Figure \ref{fig:prim_slice}.  At low densities, we see $\Pi/P_{+\nu_e} \sim 0$, indicative of no pions at these lower densities $\lesssim 0.1 n_b / n_\mathrm{sat} $.  This is supported by the middle panel of Figure \ref{fig:swan}, which also shows no charged pions in this density regime.  At higher densities, this pressure increases to a maximum of $\sim 6\%$ at $n_b \sim 2 n_\mathrm{sat}$, followed by a sharp decrease to $\Pi \lesssim 0$.  This increase corresponds to the $n_b$ profile passing through the high temperature ring.  As seen in Figure \ref{fig:temperature}, at densities many times $n_\mathrm{sat}$, the temperature drops at the center of the remnant.  In this low temperature regime, the pion condensate dominates the pion production.  Since the condensate does not contribute to pressure, one sees slightly negative $\Pi/P_{+\nu_e}$ at the highest densities, indicated by shaded regions.

The light green points in Figure \ref{fig:delta_p_vs_rho} show $\Pi/P_{+\nu_e}$ for the $+ \nu_{e,\mu} + \pi + \mu$ case.  When comparing these light green points with the darker blue points, one can quantify the influence of muons on the NS merger dynamics.  At $n_b \lesssim 0.1 n_\mathrm{sat}$, we note $\Pi/P_{+\nu_e} \sim 0.02$.  As explained in the right panel of Figure \ref{fig:prim_slice}, this lower density material is supported by thermal contributions from muonic species.  As $n_b$ increases, we see a similar shape in the $\Pi/P_{+\nu_e}$ vs $n_b$ curve to the $\nu_e + \pi$ case, systematically shifted by $\sim 2 \%$.  This behavior is expected.  Observe the top row of Figure \ref{fig:9_panel_SFHo}, focusing on the $\pi_c$ blue, dashed line.  When comparing the $+\nu_e + \pi$ case with the $+\nu_{e,\mu} + \pi + \mu$ cases, there is a decreased contribution from $\pi_c$.  This decrease in $\pi_c$ is due to the increased presence of charged muons.  Lesser amounts of pion condensate paired with more muons contributes to a higher pressure support at $n_b \gtrsim 0.4 n_\mathrm{sat}$. 

In the bottom panel, we see the same quantities for DD2.  Both variants produce similar behavior at lower densities.  However, the stiffer DD2 produces less pronounced $\Pi/P_{+\nu_e}$ beyond $n_\mathrm{sat}$. 
 Since the EOS is stiffer, it will settle to lower densities.  Furthermore, the less compressible EOS does not produce temperatures as high during the collision, compared to SFHo.  This combination of a less dense environment at lower temperatures is less conducive to creating muons and thermal pions. As such, the peak $\Pi/P_{+\nu_e}$ only reaches $\sim 0.03$.  With fewer pressure-providing pions and muons, compared to SFHo, the pion condensate allows a $\Delta P$ minimum of $\sim -0.02$.\\

\begin{figure*}
    \centering
\includegraphics[width=\textwidth]{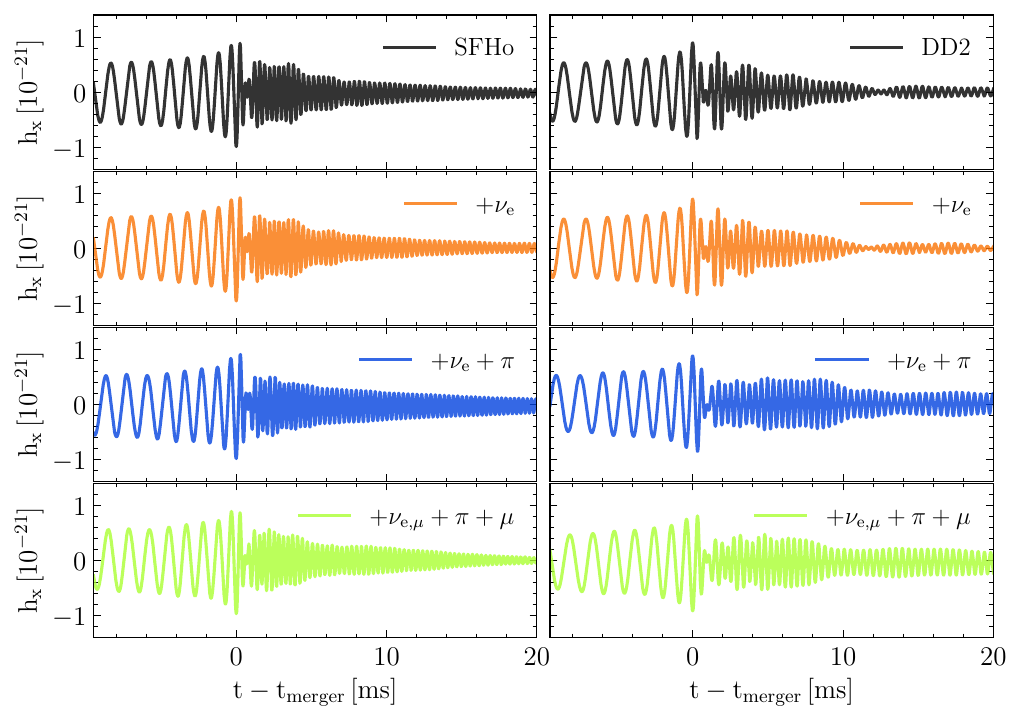}
    \caption{Gravitational wave strain, $h_\times$, for the different simulations extracted at a radius, $r\sim 1300\, \rm km$, with a detector distance of $30$ Mpc. Columns denote the two different base equations of state, DD2 and SFHo. Rows indicate the added microphysical particle content. Times, $t$, are stated relative to the time of merger, $t_{\rm merger}$.}
    \label{fig:TDWF}
\end{figure*}

\begin{figure*}
    \centering
\includegraphics[width=0.8\textwidth]{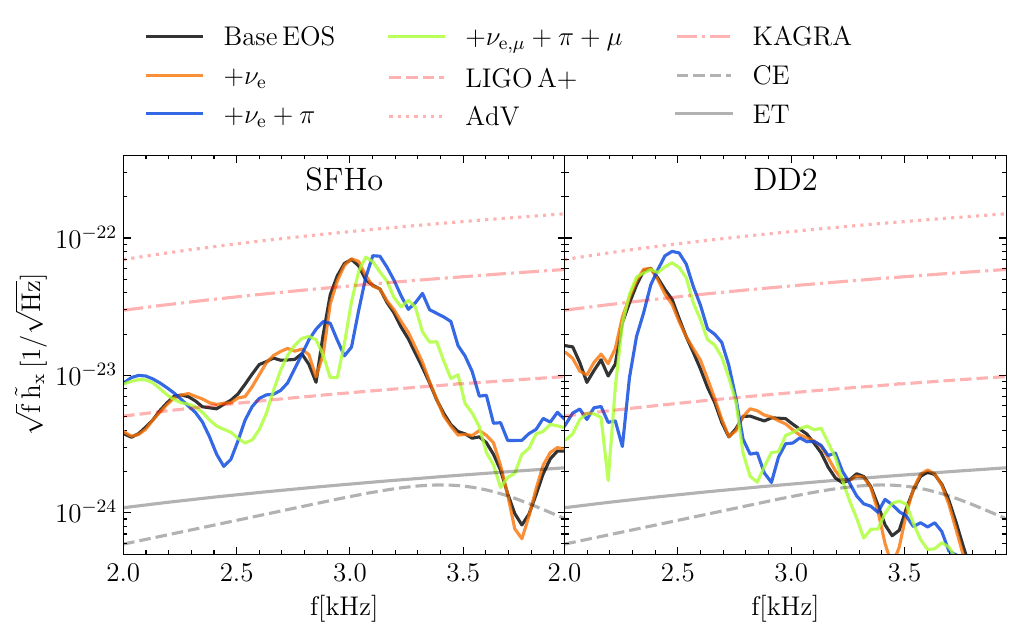}
    \caption{Post-merger gravitational wave frequency spectrum at a distance of $30\, \rm Mpc$. Shown is the effective strain, $\tilde{h}_\times$, and the frequencies, $f$, probed in the merger. Different colors denote different particle contents, different columns vary the underlying equation of state model, either DD2 or SFHo.  Overlaid are (expected) sensitivity curves for LIGO A+, Advanced Virgo (AdV), KAGRA, Cosmic Explorer (CE), and Einstein Telescope (ET). }
    \label{fig:fourier}
\end{figure*}







\subsection{Mass Ejection}



\begin{table}[]
\begin{tabular}{l|c|c}
                     & $M^\mathrm{SFHo}_\mathrm{ejecta} [M_\odot]$ & $M^\mathrm{DD2}_\mathrm{ejecta} [M_\odot]$  \\
                     \hline
Base EOS             &   $6.9\times 10^{-4}$   &   $4.8\times 10^{-4}$   \\
$+\nu_e$             &   $6.5\times 10^{-4}$   &   $4.9\times 10^{-4}$   \\
$+\nu_e + \pi$       &   $5.3\times 10^{-3}$   &   $1.9\times 10^{-2}$   \\
$+\nu_{e,\mu} + \pi + \mu$ &   $4.2\times10^{-2}$    &   $3.0\times 10^{-3}$
\end{tabular}
\caption{Ejecta masses for each simulation at $\sim 23$ ms post merger. Shown are the integrated unbound mass ejecta $M_{\rm ejecta}$ for each simulation using the geodesic criteria. Different rows denote the added particle content to the base equations of state (EOS), DD2 and SFHo. }
\label{tab:m_ejecta}
\end{table}

The potential kilonova afterglow of a NSNS event will crucially depend on the amount and composition of unbound mass ejected during merger \cite{Metzger:2016pju}. We compute the ejected mass in our simulation by integrating the mass flux,
\begin{align}
    \dot{M}_{\rm ej} = \oint_{r={\rm const}}\,{\rm d}S\, \sqrt{\gamma} \rho \frac{u^i}{u^0}\,,
\end{align}
over a sphere at radius, $r\sim 300 \, \rm km$, for $\sim 23$ ms after the merger.  We only include mass fluxes if the matter is unbound (i.e., can escape to infinity), according to the geodesic criteria, $u_0 < -1$ (e.g., \cite{Bovard:2017dfh,Foucart:2021ikp}) and Bernoulli criteria, $h u_0 < -1$ (e.g., \cite{nedora:2019}).  In Figure \ref{fig:mass_ejecta_vs_t}, we display the time evolution of the ejecta mass for both criteria.  As expected, calculations using the Bernoulli criteria produce higher ejecta mass estimates, compared to the geodesic criteria \cite{nedora:2019}.
  In Table \ref{tab:m_ejecta}, we report the values of the masses ejected for each simulation.  The central column is mass ejecta values for SFHo ($M^\mathrm{SFHo}_\mathrm{ejecta}$) while the right column is for DD2 ($M^\mathrm{DD2}_\mathrm{ejecta}$).  
  In line with \cite{Bovard:2017mvn}, our control and $+\nu_e$ EOS cases give similar order of magnitude estimates for ejecta masses $\sim 10^{-4} M_\odot$.    

To address the influence of different species for both EOSs, between the unmodified EOS and the $+\nu_e$ case, we do not see ejecta masses change more than $\sim 7$ percent.  This narrow difference is consistent with similar behavior between the base EOS and $+\nu_e$ case seen in Figure \ref{fig:temp_vs_nsat}.  When comparing to the $+\nu_e + \pi$ cases, we note more mass by an order of magnitude.  The majority of this mass is ejected after the initial tidal stream ejection.  Due to pion condensate, both EOSs become softer.  The softening of the EOS allows the NSs to become more deformed at contact and sustain multiple bounces as it finds an equilibrium.  These bounces eject material over 10s of ms and continue increasing at the end of the simulation, as seen in Figure \ref{fig:mass_ejecta_vs_t}.  The properties of the ejecta have much slower velocities $\sim 0.15 c$, narrower $Y_{e}$ distributions, and ejecta masses $\sim 10^{-3} - 10^{-2} M_\odot$ mostly in the orbital plane, reminiscent of spiral-wave winds seen in \citep{nedora:2019,Nedora:2021eoj}. 
 Furthermore, this increase in mass can be linked to the more compact NS companions.  As seen in the left panel of Figure \ref{fig:TOV}, cold TOV stars will have pion condensate in their cores, creating more compact objects.  This softening of the EOS likely will release more shock-heated ejecta.  However, similar to \cite{vijayan:2023}, we note a larger than expected impact of pions on the ejecta mass, though the numbers we find may be subject to numerical resolution effects.  We merely focus on the qualitative effect of each species shifting the ejecta mass to lower or higher values, depending on the EOS.  Lastly, the influence of muons is nonlinear, whose effect depends on EOS, causing more ejecta for the softer SFHo and less ejecta for DD2.


We now address the distribution of material within ejecta, shown in Figure \ref{fig:mejection}.  The left column contains histograms of the $Y_{e}$ of the ejected material.  The top row is for the SFHo EOS and the bottom row is for the DD2 EOS.  For both unmodified EOSs, there is a peak in $Y_{e}$ at lower values $\sim 0.075$, which is expected as the material is neutron rich.  Similar to before, we notice little difference in the ejecta between the unmodified EOS and $+\nu_e$ cases.  We caution more robust neutrino treatments beyond leakage schemes (e.g., M1) will adjust ejecta mass properties \cite{Zappa:2022rpd,ng:2024}.  The $Y_{e}$ distribution is similar to other work that has examined $1.25 \, M_\odot$ equal mass mergers with peak at $Y_{e} \sim 0.05$.  However, for the SFHo EOS, \citep{Bovard:2017mvn} show a second $Y_{e}$ peak $\sim 0.18$, whereas our results do not.  For both EOSs, we notice similar peaks in the $Y_{e}$ distributions when including pions and muons.  Note, the higher $Y_{e} \geq 0.35$ content is lower in mass by nearly an order of magnitude, compared to the peak $Y_{e}$ ejecta, and sourced from NS atmosphere shed during the inspiral phase that is swept up by the tidal ejecta. 

In the middle panels of Figure \ref{fig:mejection}, we display the entropy per baryon of the ejecta ($s$).  All distributions display a peak $s \sim 10$ $k_\mathrm{B}$ by$^{-1}$. 
 For the unmodified EOS and $+ \nu_e$ cases, both EOS show a gradual decline in the entropy profile.  By contrast, the $+\nu_e + \pi$ and $+\nu_{e,\mu} + \pi + \mu$ cases show steep fall-offs in the $s$ distribution.  This material is sourced from $\sim 5$ ms after the initial tidal tail is launched.  Since this material is not as actively shocked as the ejecta from initial contact, we expect lower entropies.  The minor bump at $s\sim 50$ $k_\mathrm{B}$ by$^{-1}$ is from the initially shedded atmosphere.
 
In the right panel, we see the velocity distribution of the ejecta.  For the similar base EOS and $+ \nu_e$ cases, we note a  distribution with velocity peak around $0.15-0.2 c$.  When compared to the cases that contain pions, we note a narrower velocity distribution, centered at a similar velocity peak.  The source of this effect is from less initial shock heating in the NSNS remnant, see Figure \ref{fig:temp_vs_nsat}.  In these cases, the orbital energy that would have been deposited into the fluid kinetic energy is instead sourced into creating pions and muons.  Moreover, similar to the entropy distribution, this ejecta is less shocked, leading to a slower moving secular drift.  This slower ejecta is in line with the findings of Ref. \cite{gieg:2024}.


Lastly, we investigate the spatial distribution of the ejecta, seen in Figure \ref{fig:ejecta_spatial_distribution}.  Each column corresponds to a base EOS, SFHo (left) and DD2 (right).  Each row corresponds to an EOS variant.  The top hemisphere shows the log of the ejecta mass $M_\mathrm{ej}$.  The bottom hemisphere shows the mass-weighted average of $Y_{e}$.   In the top row, both $+\nu_e$ variants exhibit mass ejection in the equatorial direction, indicated at latitudes $\lesssim 30^\circ$.   In the lower hemisphere, the $Y_{e}$ distributions are moderately brighter along the poles, since most of the neutron rich ejecta sent along the equator lowers the mass-averaged $Y_{e}$.

In the middle row, the $+\nu_e + \pi$ variants appear more axisymmetric, indicated by relatively smooth distributions at all longitudes. 
 As opposed to a single tidal tail that may exhibit asymmetries in the orbital plane, the later time ejecta from the pions appears to diffuse equally along all azimuthal directions.   The lower hemisphere shows a similar trend to the $+ \nu_e$ cases, where the lower $Y_{e}$, neutron rich, material is ejected near the equator.  During the inspiral, the artificial atmosphere that is ejected has a higher $Y_{e}$, compared to the non-pionic runs, contributing to the brighter pole regions.  The bottom row $+ \nu_{e, \mu} + \pi + \mu$ variants display similar distributions, compared to the $+ \nu_e + \pi$ variants.  Overall, the inclusion of pions seems to make the ejecta more axisymmetric.

\subsection{Gravitational Wave Emission}
\label{sec:GWs}


Post-merger gravitational wave emission has been shown to correlate strongly with the underlying nuclear physics \cite{Shibata:2005xz,Bauswein_2013,Takami:2014tva,Takami:2014zpa,Bernuzzi:2014owa}, especially if exotic degrees of freedom \cite{Sekiguchi:2011mc,Radice:2016rys,Most:2018eaw,Bauswein:2018bma,Prakash:2021wpz} and/or finite-temperature effects are accounted for \cite{Figura:2020fkj,Raithel:2021hye,Fields:2023bhs}.
Since the appearance of a trapped electron neutrino component inside the merger remnant alone may only cause small changes in the dynamics and gravitational wave emission \cite{Zappa:2022rpd,Most:2022yhe,Espino:2023dei}, it is interesting to ask whether the systematic addition of $\pi$ and $\mu$ change this picture. To this end we perform a systematic analysis and comparison of the gravitational wave emission of our simulations.

We begin by extracting the GW signal using the $\psi_4$ formalism \cite{Bishop:2016lgv,Reisswig:2010di}, and examining the time domain waveform (TDWF) 
for the dominant $\ell=m=2$ mode of the system in Figure \ref{fig:TDWF}. 
We can already see that there is overall good agreement in the inspiral and only small changes in the post-merger gravitational wave strain amplitude due to the addition of muons and pions. This is consistent with the expectation that most of the signal is governed by the cold equation of state, which is only marginally modified (see Sec. \ref{sec:tov}).

One of the main promising routes of obtaining additional information from the post-merger gravitational wave signal is via the dominant frequency peak of the gravitational wave strain \cite{Chatziioannou:2017ixj,Wijngaarden:2022sah,Criswell:2022ewn,Breschi:2022ens,Breschi:2023mdj}.
In Fig. \ref{fig:fourier} we show the frequency spectrum of the post-merger gravitational wave strain, focusing mainly on the dominant frequency peak. We begin by comparing the impact of adding different microphysical correction to the equation of state. In line with previous works, we find that the adding of trapped electron neutrinos does not appreciable alter the gravitational wave emission \cite{Zappa:2022rpd,Most:2022yhe,Hammond:2021vtv}, although current simulations are likely not in a convergent regime \cite{Espino:2023dei,Most:2022yhe}. We investigate the impact of adding $\pi$ to the EOS. We already found earlier that pions lead to a net softening (more compact neutron stars) of the equation of state, which generically should translate to higher frequencies. In line with this expectation, we find that for both EOS, adding pions causes a shift of around $100\,\rm Hz$ of the post-merger gravitational wave signal, although the impact on SFHo seems slightly more pronounced. 
 Such difference is consistent with uncertainties of quasi-universal relations \cite{Vretinaris:2019spn,Topolski:2023ojc}.
Finally, we also investigate the impact of $\mu$, and trapped $\nu_\mu$. Similar to the addition of trapped $\nu_e$, we find 
that the impact on the post-merger gravitational wave signal is small, though we caution that $\mu$ can in principle regulate the population of $\pi$, as seen in the amount of pion condensation (Fig. \ref{fig:slices_pions}).
 While these frequencies are out of the current frequency sensitivity bands for the LIGO-Virgo-KAGRA detectors \cite{barsotti}, these calculations stress the importance of capturing proper microphysics for accurate predictions of future NSNS waveforms to optimize future high frequency GW detectors.


\section{Conclusions}
\label{sec:conclusions}

In this work, we have investigated the impact of trapped electron and muon neutrinos, muons and pions on the post-merger phase of a neutron star coalescence. Assuming that neutrinos are trapped and that the muon and pion decay are in equilibrium, we have constructed several equations of state systematically varying the microphysics content. We have then performed fully general-relativistic neutron star merger simulations to study the impact of these equations of state on the mass ejection and gravitational wave emission. Our results also present a detailed investigation into the regions of the remnant most sensitive to pion condensation.\\  Our main findings are consistent with previous works \cite{vijayan:2023,loffredo:2023,gieg:2024}, and can be summarized as follows:\\
pions have a measurable impact on the temperature distribution of matter in the remnant from $\sim 0.01 n_\mathrm{sat} - n_\mathrm{sat}$.
When present with pions, the inclusion of muons limits the amount of pion condensate formed in the merger remnant (Figs. \ref{fig:9_panel_SFHo} and \ref{fig:9_panel_DD2}).
For matter in the core of the remnant, because $\mu_n - \mu_p$ is limited by the pion mass, $\nu_e$ and $\bar{\nu}_e$ production will be limited. In the presence of muons, thermal pions provide pressure support in the high temperature ring around the remnant center, while the muons provide pressure support in the lower density material, Figure \ref{fig:prim_slice} and Figure \ref{fig:delta_p_vs_rho}. 
For softer EOSs, in the presence of muons, the peak GW frequency can shift higher by $\sim 5 \%$.  For stiffer EOSs, this peak only changes marginally, Figure \ref{fig:fourier}. 
Overall, the frequency shifts in the post-merger gravitational wave spectrum are of comparable order to those of finite-temperature \cite{Raithel:2023zml,Fields:2023bhs} and weak-interaction effects \cite{Most:2022yhe,Espino:2023dei}.\\
Although our results are part of a first number of steps towards a consistent inclusion of muon and muon neutrinos into neutron star merger simulations, they need to be improved in several ways.
First, neutrinos should be modelled using fully consistent radiation transport schemes \cite{Foucart:2017mbt,Radice:2021jtw,Izquierdo:2023fub,ng:2024}. This is particularly important in correctly tracking the muonic lepton fraction, which we have crudely approximated here. Furthermore, we treat $\pi$ as ideal Bose-Einstein gasses, while treating $\mu$ as ideal Fermi gasses.  To be more consistent, modeling the transport of these charged particles would be ideal.  We also neglect the density dependence of the pion mass.  Similar to \cite{vijayan:2023}, we use the vacuum mass to present optimistic estimates, to see if these effects are observable under ideal conditions. \\
For future work, we plan to also extend beyond the NSNS system and better quantify the influence of these particles in the core-collapse supernova context \cite{Bollig:2017lki}.  Furthermore, how $\pi + \mu$ may couple with more exotic particles like axions remains an interesting route to pursue \cite{Diamond:2023cto,dev:2024,Manzari:2024jns}. 

\begin{acknowledgments}
The authors are grateful for helpful conversations with Eleonora Loffredo, Ninoy Rahman, and Javier Roulet.  Analysis for this work was done using \texttt{matplotlib} \cite{Hunter:2007}, \texttt{numpy} \cite{harris:2020}, \texttt{scipy} \cite{virtanen:2020} and \texttt{kuibit} \cite{bozzola:2021}.  M.A.P. was supported by the Sherman Fairchild Foundation, NSF grant PHY-2309211, PHY-2309231, and OAC-2209656 at Caltech.  ERM acknowledges supported by the National Science Foundation under grants No. PHY-2309210 and OAC-2103680.  This work mainly used Delta at the National Center for Supercomputing Applications (NCSA) through allocation PHY210074 from the Advanced Cyberinfrastructure Coordination Ecosystem: Services \& Support (ACCESS) program, which is supported by National Science Foundation grants \#2138259, \#2138286, \#2138307, \#2137603, and \#2138296.  Additional simulations were performed on the NSF Frontera supercomputer under grant AST21006.
\end{acknowledgments}

\appendix
\section{Repopulating the Equation of State with New Species}
\label{app:eos_steps}

\subsection{The Choice of Muonic Lepton Fraction $Y_{l,\mu}$}
\label{sec:ylmu_cold}
The choice of $Y_{l,\mu}$ requires some care.  Informed by the results of \cite{loffredo:2023}, the muonic lepton fraction at densities below $10^{14}$ g cm$^{-3}$ has been shown to be trapped at small values. Following Appendix A in \cite{vijayan:2023}---and similar to the value used in \cite{loffredo:2023}---we fix $Y_{l,\mu} = 0.01$, which has been shown to replicate muon lepton fractions in merger simulations.  However, as the field evolves, more accurate approximations for $Y_{l,\mu}$ exist \cite{gieg:2024,ng:2024}.

Above the density cut of $10^{14}$ g cm$^{-3}$, the NS matter has been shown to trap the local muonic particle fractions of the cold, isolated NS companions before merger \cite{loffredo:2023}.  Hence, our goal is to construct an expression of muonic lepton fraction as a function of density $\tilde{Y}_{l,\mu}(\rho)$.  Begin by assuming cold, neutrinoless $\beta$ equilibrium.  By cold, we select a temperature of 0.1 MeV that is near the bottom of the EOS tables.  Equation (\ref{eq:chem}) provides two expressions for $\beta$ equilibrium and its relation to the muonic chemical potential: 
\begin{equation}
    \mu_e(\rho, T_\mathrm{low}, Y_p) = \mu_n(\rho, T_\mathrm{low}, Y_p) -  
    \mu_p(\rho, T_\mathrm{low}, Y_p),
\end{equation}
\begin{equation}
    \mu_\mu(\rho, T_\mathrm{low}, Y_p) = \mu_e(\rho, T_\mathrm{low}, Y_p).
    \label{eq:mu_equil}
\end{equation}
To construct $\tilde{Y}_{l,\mu}(\rho)$ using this two equation system, we use an iterative procedure.  
\begin{enumerate}
\item At a given $\rho$, begin with a trial guess $\hat{Y}_p$. 

\item From the old EOS, use $(\rho, T_\mathrm{low}, \hat{Y}_p)$ to find $\mu_e(\rho, T_\mathrm{low}, \hat{Y}_p)$.

\item Using Equation (\ref{eq:n_fermi}), $T_\mathrm{low}$, and $\mu_e$, calculate $n_{e^\pm}$.

\item Convert to $Y_e = (n_{e^-} - n_{e^+}) / (\rho / m_\mathrm{baryon})$.

\item  Using Equation (\ref{eq:mu_equil}), equate $\mu_\mu = \mu_e(\rho, T_\mathrm{low}, \hat{Y}_p)$.

\item Using Equation (\ref{eq:n_fermi}), $T_\mathrm{low}$, and $\mu_\mu$, calculate $n_{\mu^\pm}$.

\item Convert to $Y_\mu = (n_{\mu^-} - n_{\mu^+}) / (\rho / m_\mathrm{baryon})$.

\item Using Equation (\ref{eq:charge}), update $\hat{Y}_p = Y_e + Y_\mu$.

\item If $\hat{Y}_p$ does not converge, return to step 2. 
 If $\hat{Y}_p$ converges, record $Y_\mu$, move to the next $\rho$, and return to step 1.
\end{enumerate}
Condensing our assumptions regarding the muons

\begin{equation}
    Y_{l,\mu}(\rho) =
    \begin{cases}
    & \tilde{Y}_{l,\mu}(\rho), \, \mathrm{if} \, \rho \geq 10^{14} \mathrm{g\,} \mathrm{cm}^{-3}\\
    & 0.01, \, \mathrm{if} \, \rho < 10^{14} \mathrm{g\,} \mathrm{cm}^{-3}.
    \end{cases}
    \label{eq:ymu_cold}
\end{equation}

\subsection{No Pion Condensate}
\label{sec:no_condensate}
 At each entry of EOS$_\mathrm{old} (\rho, T, Y_p)$, baryon chemical potentials $(\mu_n - \mu_p)$ are provided.  If $(\mu_n - \mu_p) < m_\pi^-$, a pion condensate will not form.  The unmodified EOS table provides thermodynamic quantities in terms of charge fraction EOS$_\mathrm{old}(\rho, T, Y_p)$.  We retabulate in terms of lepton fraction of electrons, $Y_{l,e}$; thus the first term on the right-hand side of Equation (\ref{eq:charge_trial}) is fixed at a given thermodynamic entry $(\rho, T, Y_{l,e})$.  At a given $\rho$, $T$, $Y_{l,e}$ begin with a guess of $\hat{Y}_p = Y_{l,e}$; calculate $\mu_n$, $\mu_p$, and $\mu_e$ from EOS$_\mathrm{old}(\rho, T, \hat{Y}_p)$.  Noting chemical equilibrium,
\begin{equation}
    \mu_{\nu_e} = \mu_e(\rho,T,\hat{Y}_p) - \mu_n(\rho,T,\hat{Y}_p) + \mu_p(\rho,T,\hat{Y}_p).
\end{equation}
We calculate the  $(Y_{\nu_e}(\mu_{\nu_e}, \rho, T) - Y_{\bar{\nu}_e}(\mu_{\nu_e}, \rho, T))$ terms using Equation (\ref{eq:nu_number}).  With a known $\rho$, use Equation (\ref{eq:ymu_cold}) to calculate $Y_{l,\mu}$.

To calculate the $(Y_{\nu_\mu} - Y_{\bar{\nu}_\mu})$ terms, we remind ourselves of Equation (\ref{eq:mu_lepton}), with explicit dependencies

\begin{equation}
    Y_{l,\mu} = Y_\mu (\hat{\mu}_\mu, \rho, T) + Y_{\nu_\mu} (\hat{\mu}_{\nu_\mu}, \rho, T) - Y_{\bar{\nu}_\mu} (-\hat{\mu}_{\nu_\mu}, \rho, T).
    \label{eq:mu_lepton_dependence}
\end{equation}
In Equation (\ref{eq:mu_lepton_dependence}), $Y_{l,\mu}$ is fixed according to Equation (\ref{eq:ymu_cold}).  Using Equation (\ref{eq:chem}), $\hat{\mu}_{\nu_\mu}$ can be written as
\begin{equation}
    \hat{\mu}_\mu = \mu_n(\rho, T, \hat{Y}_p) - \mu_p(\rho, T, \hat{Y}_p) + \hat{\mu}_{\nu_\mu}.
    \label{eq:mu_lepton_chem_potential}
\end{equation}
Applying Equation (\ref{eq:n_fermi}) and Equation (\ref{eq:nu_number}) to Equation (\ref{eq:mu_lepton_dependence}), one can solve for $\hat{\mu}_{\nu_\mu}$ numerically, as $\hat{\mu}_{\nu_\mu}$ is the only unknown in the equation.  Using the newly found $\hat{\mu}_{\nu_\mu}$, Equation (\ref{eq:nu_number}) gives the needed $Y_{\nu_\mu} - Y_{\bar{\nu}_\mu}$ in Equation (\ref{eq:charge_trial}). 

For $Y_\pi$, note Equation (\ref{eq:charge}) to obtain $\mu_\pi = \mu_n(\rho, T, \hat{Y}_p) - \mu_p(\rho, T, \hat{Y}_p)$.  Apply Equation (\ref{eq:n_bose}) to yield $Y_\pi$.  Adding together all particle fractions gives a new $\hat{Y}_p$.  Return to the beginning of the procedure to repeat until convergence.  Once a converged $Y_{p,\mathrm{new}}$ is obtained, move to the next $(\rho, T, Y_{l,e})$ entry and repeat.

\subsection{Pion Condensate}
\label{sec:pion_condensate}

In certain thermodynamic settings, $(\mu_n - \mu_p) > m_\pi^-$, a pion condensate can form in the NS remnant.  In this case, the procedure to calculate $Y_p$ changes.  Begin with $(\rho, T, Y_{l,e})$.  Select a trial $\hat{Y}_p = Y_{l,e}$.  Repeat the procedure in Section \ref{sec:no_condensate} until convergence, giving a temporary $Y_{p\mathrm{,temp}}$.  If EOS$_\mathrm{old}(\rho, T, Y_{p \mathrm{,temp}})$ produces $(\mu_n - \mu_p) > m_\pi^-$, a condensate will be produced. 
 In this case, as in Section II of \cite{vijayan:2023}, begin by iterating over EOS$_\mathrm{old}$ to find $Y_p$ for 
\begin{equation}
  \mu_n(\rho, T, \hat{Y}_p) - \mu_p(\rho, T, \hat{Y}_p) = m_{\pi^-},
\end{equation}
the maximum chemical potential for pions.  With $Y_p$ in hand, observe the governing charge neutrality condition when condensate is present
\begin{equation}
    Y^\mathrm{cold}_\pi = Y_p - Y_e - Y_\mu - Y^\mathrm{thermal}_\pi.
\end{equation}
Expanding with the definition of lepton fractions gives,
\begin{equation}
    Y^\mathrm{cold}_\pi = Y_p - Y_{l,e} + (Y_{\nu_e} - Y_{\bar{\nu}_e}) + Y_{l,\mu} + (Y_{\nu_\mu} - Y_{\bar{\nu}_\mu}) + Y^\mathrm{thermal}_\pi.
    \label{eq:condensate_expanded}
\end{equation}
As before, $Y_{l,e}$ is the known EOS entry.  With a known $(\rho, T, Y_p)$ one calculates $\mu_n$, $\mu_p$, and $\mu_e$, and, using Equation (\ref{eq:chem}), calculates $\mu_{\nu_e}$.  Applying Equation (\ref{eq:nu_number}) gives $Y_{\nu_e} - Y_{\bar{\nu}_e}$. As density is known, $Y_{l,\mu}$ is given by Equation (\ref{eq:ymu_cold}).  As before, a similar iterative procedure is followed to solve Equation (\ref{eq:mu_lepton_dependence}) when $\mu_{\mu} = \mu_{\nu_\mu} + m_{\pi^-}$, giving $Y_{\nu_\mu} - Y_{\bar{\nu}_\mu}$.  Lastly, $Y^\mathrm{thermal}_\pi$ is calculated using Equation (\ref{eq:n_bose}) for $\mu_{\pi^\pm} = \mp m_{\pi^-}$.  Any remaining charge after evaluating Equation (\ref{eq:condensate_expanded}) represents the charge fraction of the condensate.  In certain rare cases near the edge of the EOS table, and at temperatures beyond realistic simulation temperatures $\gtrsim 100$ MeV, the right-hand side of Equation (\ref{eq:condensate_expanded}) can produce a negative number.  In these cases, we floor the value of $Y_\pi^c = 0$.


While in certain proton rich conditions $\mu_{\pi^+}$ can exceed the charged pion mass, we do not expect these conditions in NSNS mergers \cite{vijayan:2023}, so we do not account for the effect of positively charged pion condensate.

\subsection{Retabulating the EOS}

At this point, the chemical potentials and particle fractions have been calculated for the new species.  With these updated species, the charge, or proton, fraction has now changed. Our original EOS is tabulated according to rest mass density, temperature, and proton (charge) fraction $Y_p$ as $(\rho, T, Y_p)$.  With the addition of pions and muons, this brings the total number of charge carriers to four: protons, electrons/positrons, muons, and pions.  In principle, this means thermodynamic quantities are uniquely determined by a six dimensional parameter space $(\rho, T, Y_p, Y_e, Y_\mu, Y_\pi)$.  Similar to the unmodified EOS, charge neutrality in Equation (\ref{eq:charge}) brings it down to five dimensions $(\rho, T, Y_p, Y_\mu, Y_\pi)$.  Assuming a known prescription for $Y_{l,\mu}$ brings the unique degrees of freedom to $(\rho, T, Y_p, Y_\pi)$ and assuming chemical equilibrium through Equation (\ref{eq:chem}) reduces the set to three thermodynamic dimensions $(\rho, T, Y_{p})$.

In our simulations, one of our assumptions is that the lepton number is advected along with the fluid, consistent with \cite{loffredo:2023}.  Thus, to match the EOS calls within \texttt{FIL}, we must now retabulate the EOS according to $Y_{l,e}$ instead of $Y_p$; essentially an EOS call will go from EOS$_\mathrm{old}$($\rho$, T, $Y_{p,\mathrm{old}} = Y_e$) to EOS$_\mathrm{new}$($\rho$, T, $Y_{l,e}$).  In words, the third thermodynamic index now will represent $Y_{l,e}$, not $Y_p$. Originally, quantities like the sound speed, baryon chemical potentials, and pressure were uniquely determined by ($\rho$, T, $Y_{p,\mathrm{old}} = Y_e$).  However, with the updated charge fraction, we loop through thermodynamic space ($\rho$, T, $Y_{l,e}$), remembering every entry has a corresponding $Y_{p,\mathrm{new}}$.  For every ($\rho$, T, $Y_{l,e}$) entry, set EOS$_\mathrm{new}$($\rho$, T, $Y_{l,e}$) = EOS$_\mathrm{old}$($\rho$, T, $Y_{p,\mathrm{new}}$).  This equation signifies updating already existing quantities like the sound speed, baryon chemical potentials, and pressure to be consistent with the new proton fractions.  As a second step, we must add contributions to the pressure and specific internal energy, now that new particle species are present.  This returns the discussion to Section \ref{sec:thermo_qts}.

    

\bibliography{main,inspire}

\end{document}